\documentclass[aps,prl,amsmath,amssymb,twocolumn,showpacs,superscriptaddress,groupedaddress,nofootinbib]{revtex4-1}

\usepackage{graphicx,amssymb,amsmath}
\usepackage[dvips]{xcolor}
\usepackage{hyperref}
\usepackage{amsmath}
\usepackage{amssymb,gensymb}
\newcommand{\be}{\begin{equation}}
\newcommand{\ee}{\end{equation}}

\begin{document}
\title{ 
Analysis of the  MINERvA antineutrino double-differential cross sections within the SuSAv2-MEC model 
}

 \author{G.D.~Megias}
 \affiliation{Departamento de F\'{i}sica At\'omica, Molecular y Nuclear, Universidad de Sevilla, 41080 Sevilla, Spain}
\affiliation{IRFU, CEA, Universit\'e Paris-Saclay, 91191 Gif-sur-Yvette, France}
 \author{M.B.~Barbaro}
 \affiliation{Dipartimento di Fisica, Universit\`{a} di Torino and INFN, Sezione di Torino, Via P. Giuria 1, 10125 Torino, Italy}
 \author{J.A.~Caballero}
 \affiliation{Departamento de F\'{i}sica At\'omica, Molecular y Nuclear, Universidad de Sevilla, 41080 Sevilla, Spain}

 \author{S.~Dolan}
 \affiliation{Ecole Polytechnique, IN2P3-CNRS, Laboratoire Leprince-Ringuet, 91120 Palaiseau, France}
\affiliation{IRFU, CEA, Universit\'e Paris-Saclay, 91191 Gif-sur-Yvette, France} 

\date{\today}
\begin{abstract}
We compare the results of the SuSAv2 model including meson-exchange currents (MEC) with the recent measurement of the quasielastic-like double differential antineutrino cross section on hydrocarbon (CH) performed by the MINERvA  Collaboration~\cite{Patrick:2018gvi}. The relativistic nature of the model makes it suitable to describe these data, which correspond to a mean beam energy of 3.5 GeV.
The standard SuSAv2 model predictions agree well with the data without needing any additional or tuned parameter. The role of longitudinal MEC is non-negligible and  improves the agreement with the data. 
We also consider the impact of different treatments of the $\Delta$-resonance propagator in the two-body currents on the data comparison.
\end{abstract}


\maketitle

Neutrino oscillation physics has become a very active field of study in recent years.
The knowledge and understanding of neutrino interactions with nucleons and nuclei is a basic requirement in order to provide neutrino properties, i.e., oscillation parameters, including the CP violating phase, and the neutrino mass differences with high accuracy. The intense experimental activity witnessed in recent years, and planned in the near future, with several long baseline neutrino experiments 
making use of complex nuclear targets, aims to measure basic neutrino properties with unprecedented precision. However, this can only be accomplished by having an excellent control of the medium effects in neutrino-nucleus scattering, which represent one of the most important sources of systematic uncertainty in the experimental analyses~\cite{Alvarez-Ruso:2017oui}.

Most of the recent (MiniBooNE, T2K, MINERvA, NOvA) and future (DUNE, HyperK) neutrino experiments cover a wide neutrino energy range; the neutrino fluxes can extend from hundreds of MeV to several GeV. This can imply very different values of the energy and momentum transfers, hence requiring in some of the cases a realistic description of different reaction mechanisms: quasielastic, two-particle-two-hole (2p2h) meson exchange currents, nucleon resonances, pion production, inelastic processes, etc. Furthermore, the large energy and momentum values involved in most of the experiments make it necessary to incorporate relativity as an essential ingredient in the process; not only to describe properly the weak reaction mechanism, but also the nuclear dynamics~\cite{Alvarez-Ruso:2017oui,Katori:2016yel,Alvarez-Ruso:2014bla}.
  
The SuSAv2 model~\cite{Gonzalez-Jimenez:2014eqa} is an improved version of the Super-Scaling Approach introduced in \cite{Amaro:2004bs}, which exploits the scaling and superscaling~\cite{Donnelly:1998xg} properties of inclusive electron scattering data in order to predict neutrino-nucleus observables.
The model is fully relativistic and takes into account the behaviour of the nuclear responses predicted by the Relativistic Mean Field (RMF): in particular, the natural enhancement of the transverse electromagnetic response, a genuine dynamical relativistic effect, is incorporated in the model. Moreover, at high momentum transfer, where the RMF fails due to the strong energy-independent relativistic potentials, the SuSAv2 model incorporates a smooth transition to the relativistic plane wave impulse approximation (RPWIA), more appropriate to describe the high $q$ domain. The parameters associated to the RMF/RPWIA mixing are fixed once and for all by fitting the high quality $(e,e')$ data on different nuclei~\cite{Megias:2016lke}. The model also includes ingredients beyond the impulse approximation, namely 2p2h excitations. These contributions, corresponding to the coupling of the probe to a pair of interacting nucleons and associated to two-body meson exchange currents (MEC)~\cite{Amaro:2010sd,Amaro:2011aa,RuizSimo:2016ikw}, are known to play a very significant role in the ``dip" region between the quasielastic (QE) and $\Delta$ peaks. In the SuSAv2 approach they are treated within the Relativistic Fermi Gas (RFG) model, which allows for a fully relativistic calculation, as required for the extended kinematics involved in neutrino reactions. It is also worth mentioning that the present SuSAv2-MEC model in its present form can only predict inclusive reactions. Work is in progress to extend the semi-inclusive prediction of the RMF 
for $(e,e'p)$ reactions~\cite{PhysRevC.64.024614} to the neutrino case as well as to produce 2p2h semi-inclusive results.

The SuSAv2 calculation of the inclusive $(e,e')$ cross section on $^{12}$C, presented in \cite{Megias:2016lke}, provides a remarkably good description of the data for very different kinematical situations. In order to perform such comparisons the SuSAv2 model has been extended from the quasielastic domain to the inelastic region by employing phenomenological fits~\cite{Bosted1,Bosted2} to the single-nucleon inelastic electromagnetic structure functions together with the information about the nuclear dynamics extracted from the SuperScaling Approach, as described in previous works~\cite{Barbaro:2003ie,Megias:2016lke,Megias:2017PhD}.
Comparisons of the SuSAv2 model predictions to charged-current neutrino scattering observables have been shown in \cite{Megias:2016fjk}. In particular, a good agreement has been achieved with the double differential neutrino and antineutrino charged-current quasielastic (CCQE) cross sections measured by the MiniBooNE~\cite{AguilarArevalo:2010zc} and T2K~\cite{Abe:2016tmq} experiments, which correspond to similar mean (anti)neutrino energies $<E>\sim$ 0.7-0.8 GeV, and also with former MINERvA measurements~\cite{Wolcott:2015hda,McFarlandPrivate}. The MINERvA experiment covers a higher energy range (1.5 $-$ 15 GeV), which is of great interest for the future DUNE facility~\cite{Acciarri:2015uup}. In this energy region relativistic models are needed to describe the nuclear dynamics, while non-relativistic calculations are bound to fail. In this letter we compare the SuSAv2 results with the recent MINERvA measurement of double differential antineutrino cross section on a hydrocarbon target~\cite{Patrick:2018gvi}.

In Fig.~1 we show the  double differential cross section of muonic antineutrino on hydrocarbon as a function of the transverse (with respect to the antineutrino beam) momentum of the outgoing muon, in bins of the muon longitudinal momentum. 
For the data we use the same nomenclature employed in the experimental paper~\cite{Patrick:2018gvi}. The ``QE-like" experimental points include, besides pure quasielastic contributions, events that have post-FSI final states without mesons, prompt photons above nuclear de-excitation energies, heavy baryons, or protons above the
proton tracking kinetic energy threshold of 120 MeV, thus including zero-meson final states arising from resonant pion production followed by pion absorption in the nucleus and from interactions on multinucleon states. This is similar to the so-called CC0$\pi$ definitions used by other experiments~\cite{AguilarArevalo:2010zc,Abe:2016tmq}. On the contrary, the ``CCQE" signal (also defined in other experiments as ``CCQE-like'') corresponds to events initially  generated in the GENIE neutrino interaction event generator~\cite{Andreopoulos:2009rq} as  quasi-elastic  (that is, no resonant or deep  inelastic scatters, but including scatters from nucleons in correlated pairs with zero-meson final states), regardless of the final-state particles produced, thus including CCQE and 2p2h interactions.
The difference between the two data sets, mainly due to pion production plus re-absorption, varies between $\sim15\%$ and $\sim5\%$ depending on the kinematics. According to MINERvA's acceptance, the muon scattering angle is limited to $\theta_\mu<$ 20$^{\circ}$ as well as the muon kinematics (1.5 GeV $< p_{||} < $ 15 GeV, $p_T<$ 1.5 GeV) in both experimental and theoretical results, leading to a significant phase-space restriction for large energy and momentum transfer to the nuclear target.

The theoretical curves correspond to the aforementioned SuSAv2 model and include 2p2h excitations induced by meson exchange currents. The antineutrino hydrogen contribution in the cross sections only enters through the 1p1h channel and has been evaluated by computing the elastic antineutrino-proton cross section. The present calculation does not include processes corresponding to pion emission followed by re-absorption inside the nucleus. Therefore the curves are meant to be compared with the ``CCQE" data rather than with the ``QE-like" ones. However, we also display the QE-like cross sections, to illustrate MINERvA's estimation of the magnitude of the QE-like resonance component among other minor effects.

\begin{figure*}[!h]\vspace{-0.728cm}
		\hspace*{-0.295cm}\includegraphics[scale=0.192, angle=270]{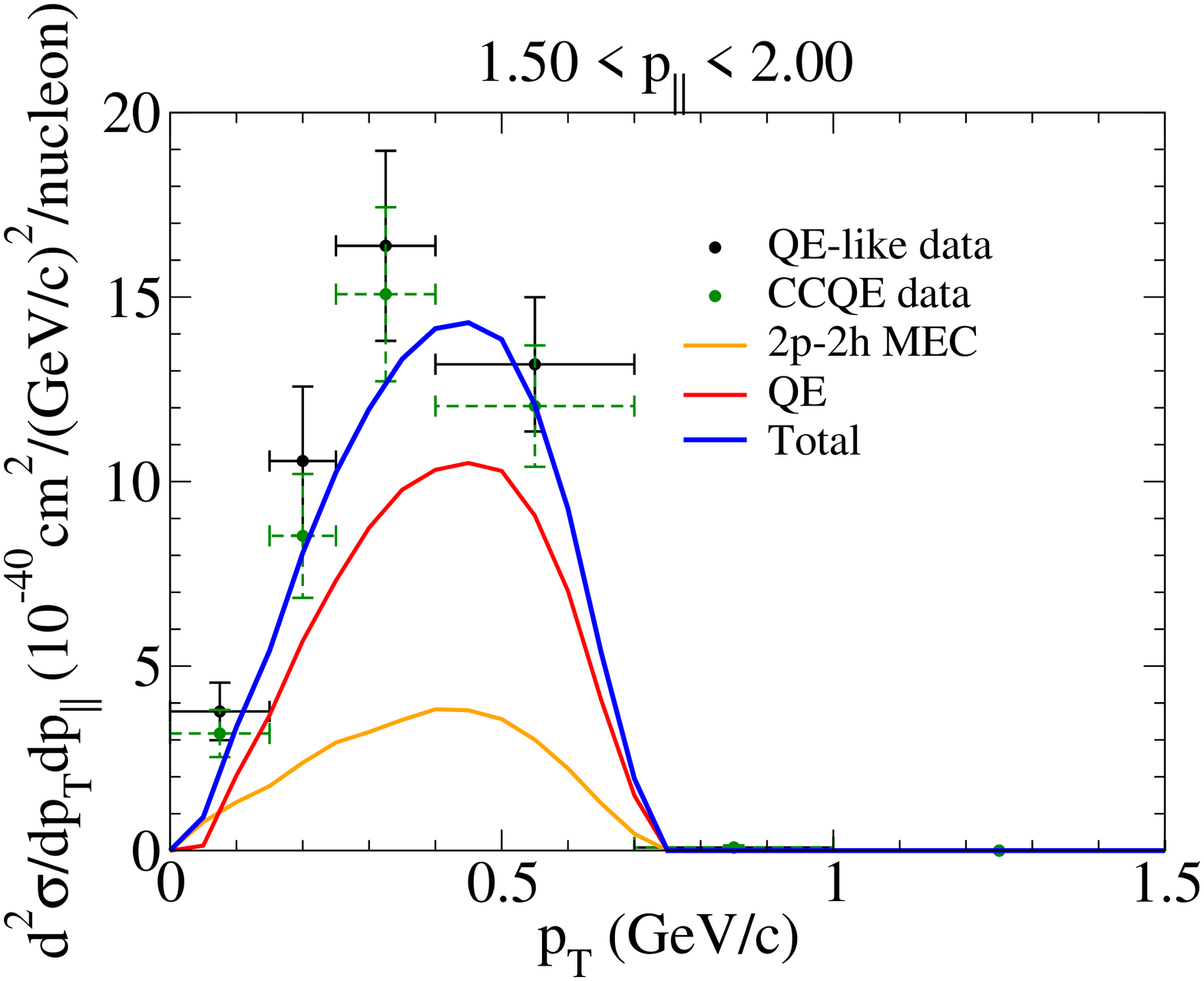}\hspace*{-0.584cm}%
		\hspace*{-0.295cm}\includegraphics[scale=0.192, angle=270]{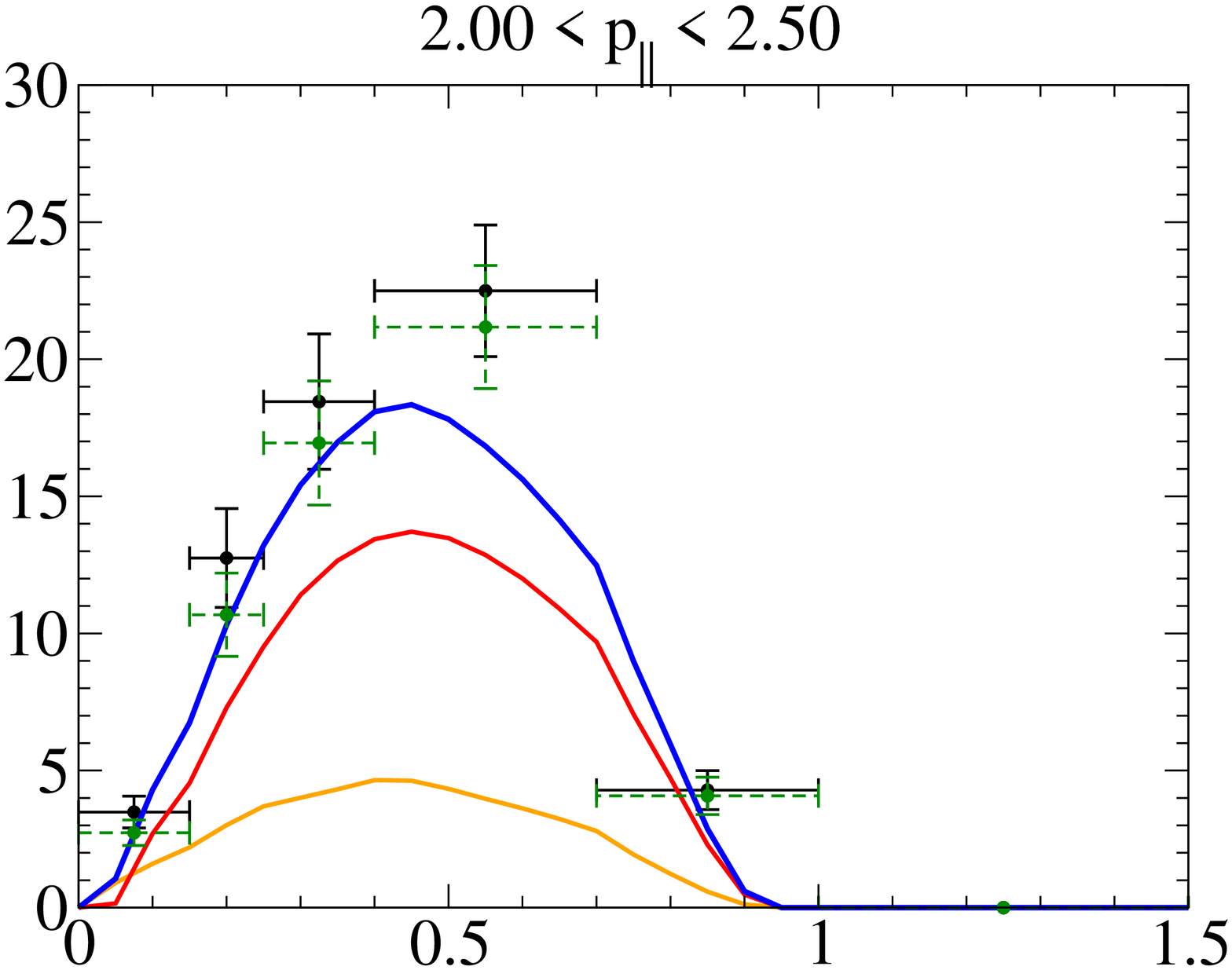}\hspace*{-0.584cm}\\%
		\hspace*{-0.295cm}\includegraphics[scale=0.192, angle=270]{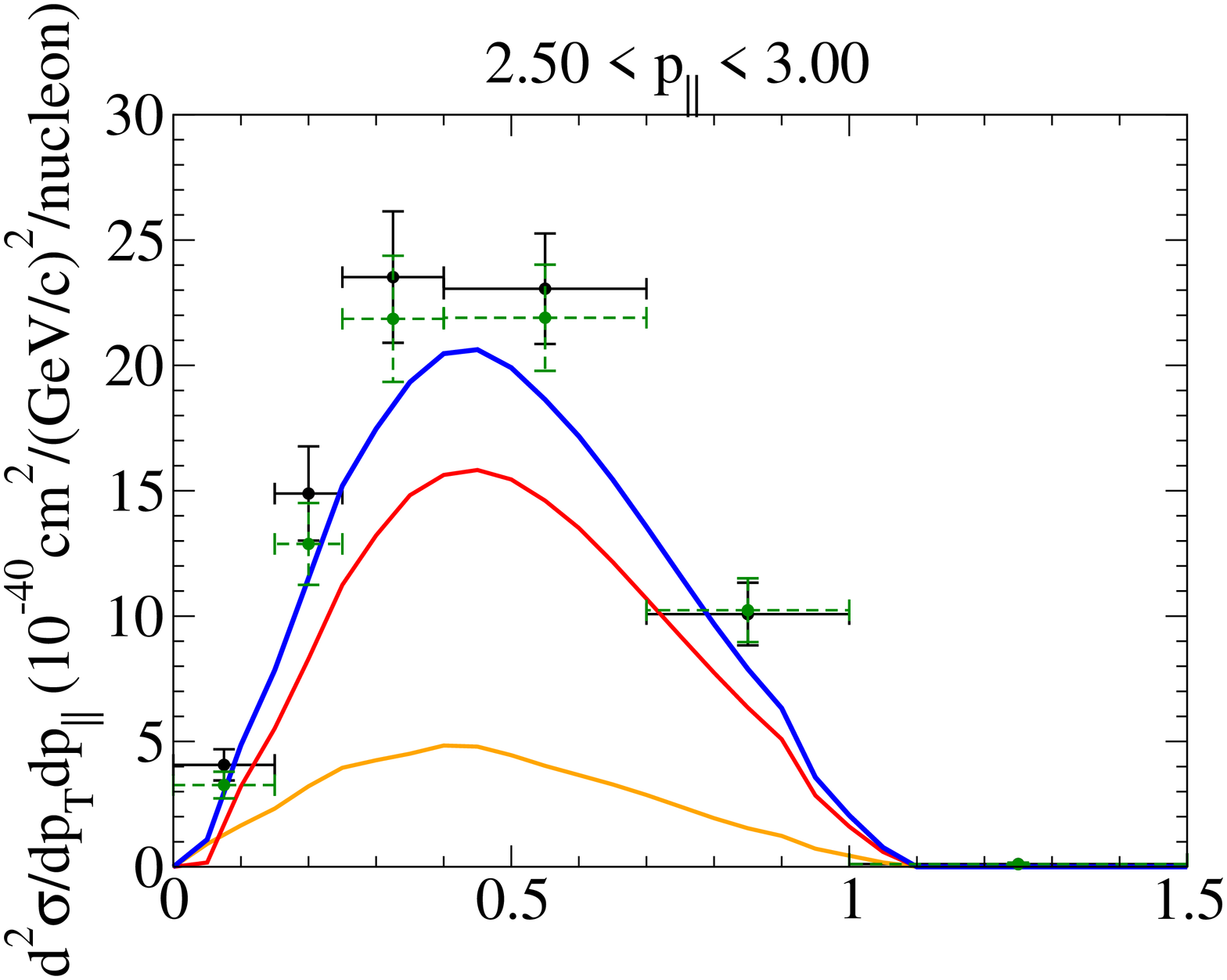}\hspace*{-0.584cm}%
		\hspace*{-0.295cm}\includegraphics[scale=0.192, angle=270]{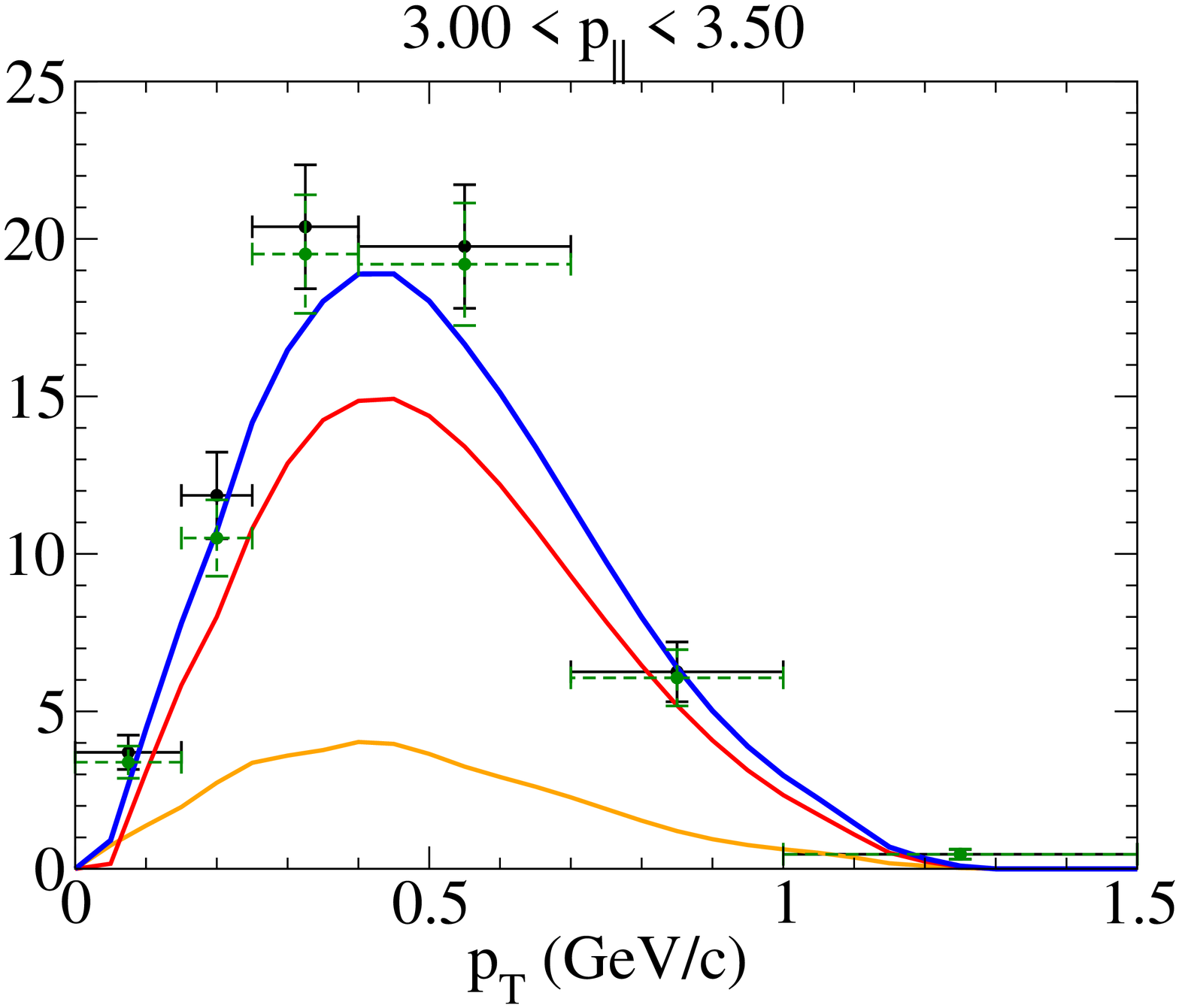}\hspace*{-0.584cm}%
		\hspace*{-0.295cm}\includegraphics[scale=0.192, angle=270]{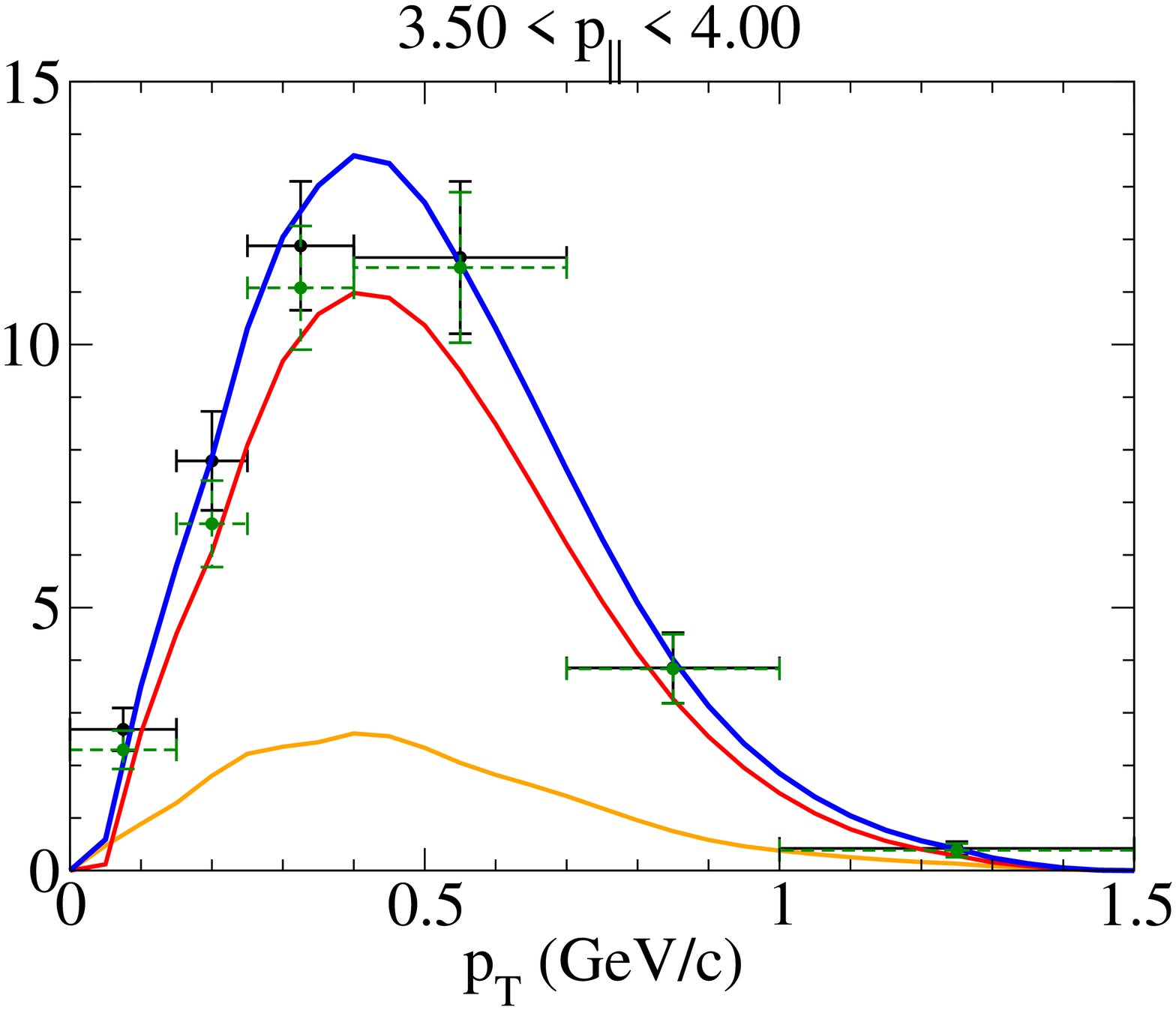}\hspace*{-0.584cm}\\%
		\hspace*{-0.295cm}\includegraphics[scale=0.192, angle=270]{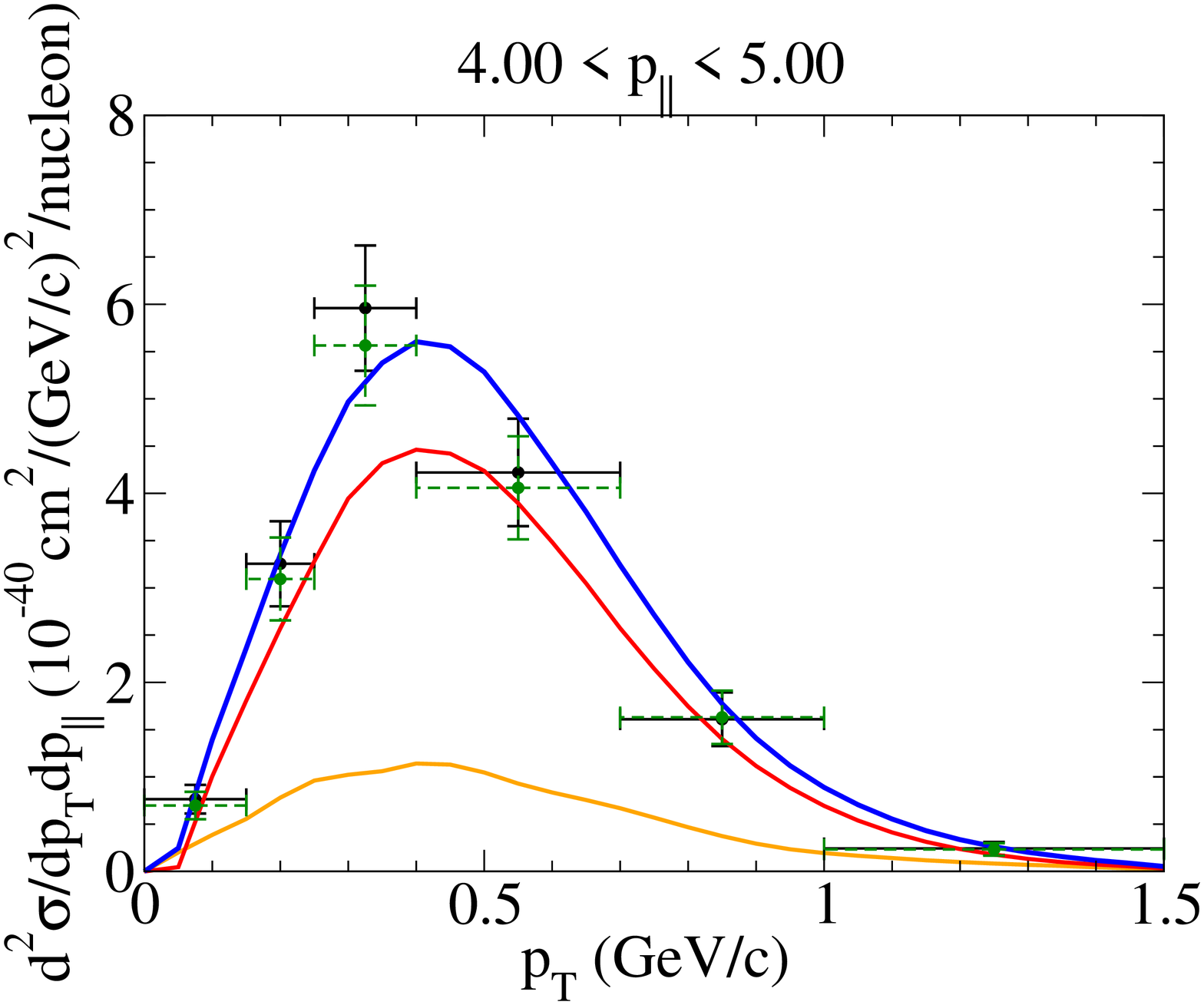}\hspace*{-0.584cm}%
		\hspace*{-0.295cm}\includegraphics[scale=0.192, angle=270]{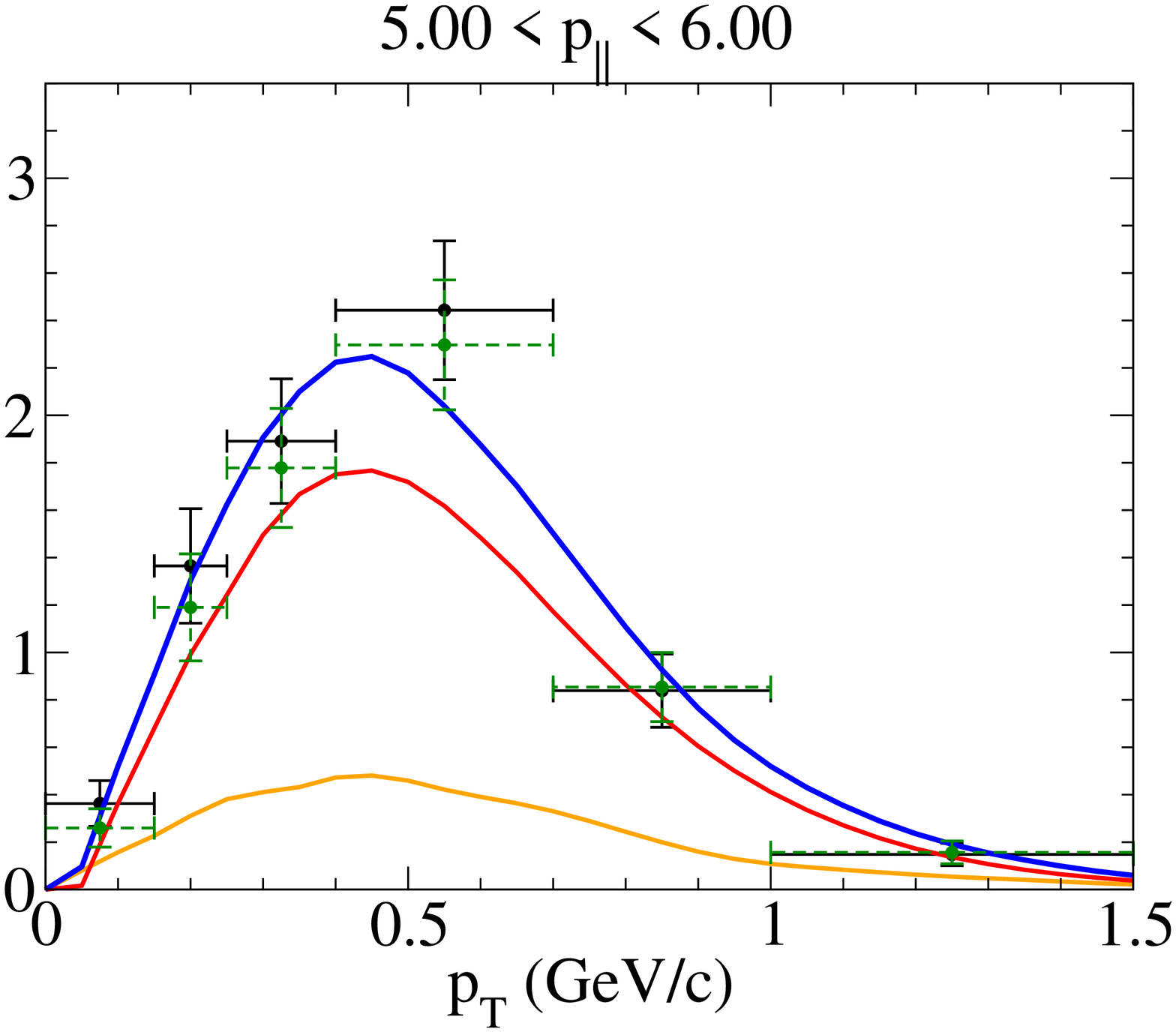}\hspace*{-0.584cm}%
		\hspace*{-0.295cm}\includegraphics[scale=0.192, angle=270]{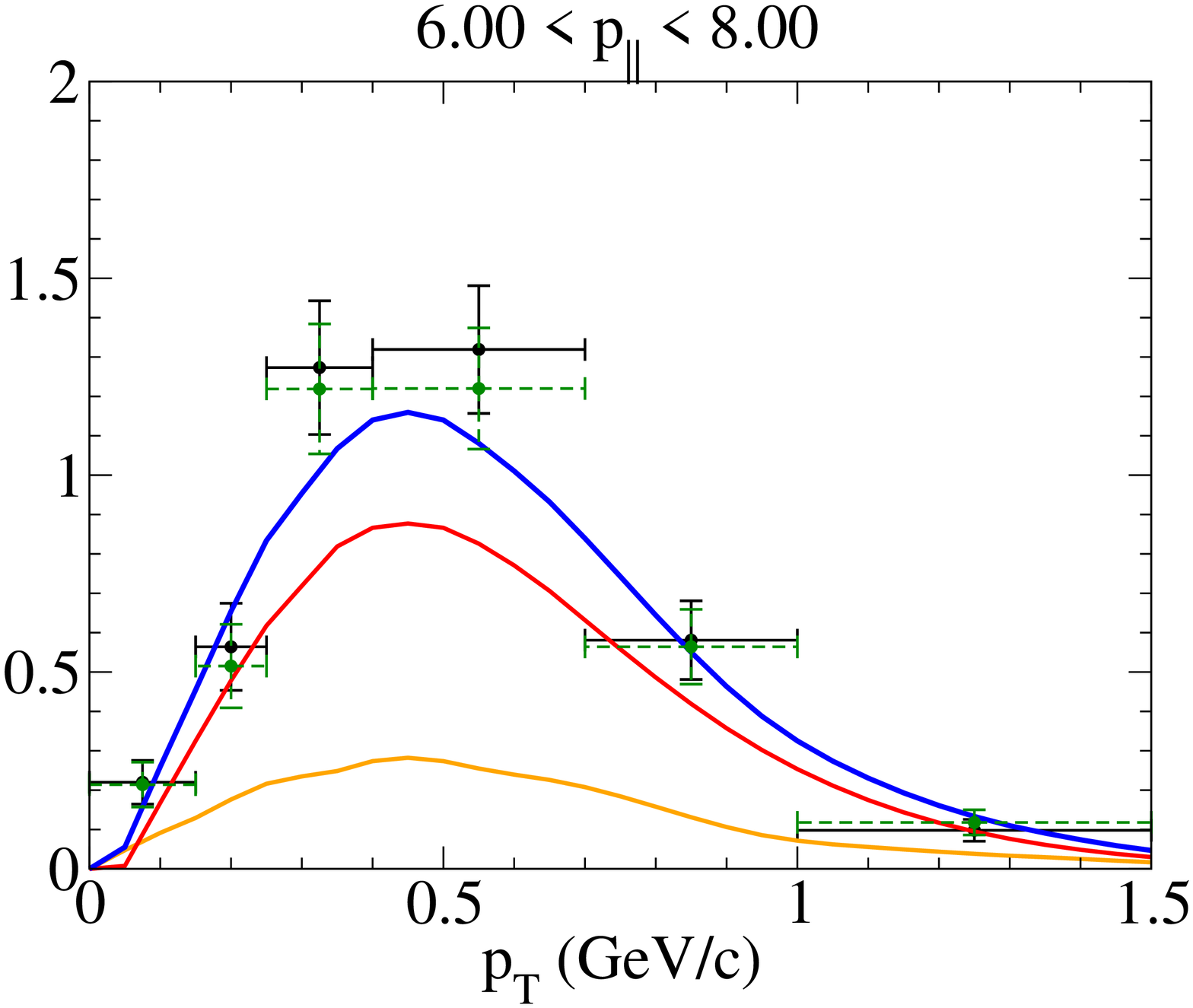}\hspace*{-0.584cm}\\%
		\hspace*{-0.695cm}\includegraphics[scale=0.192, angle=270]{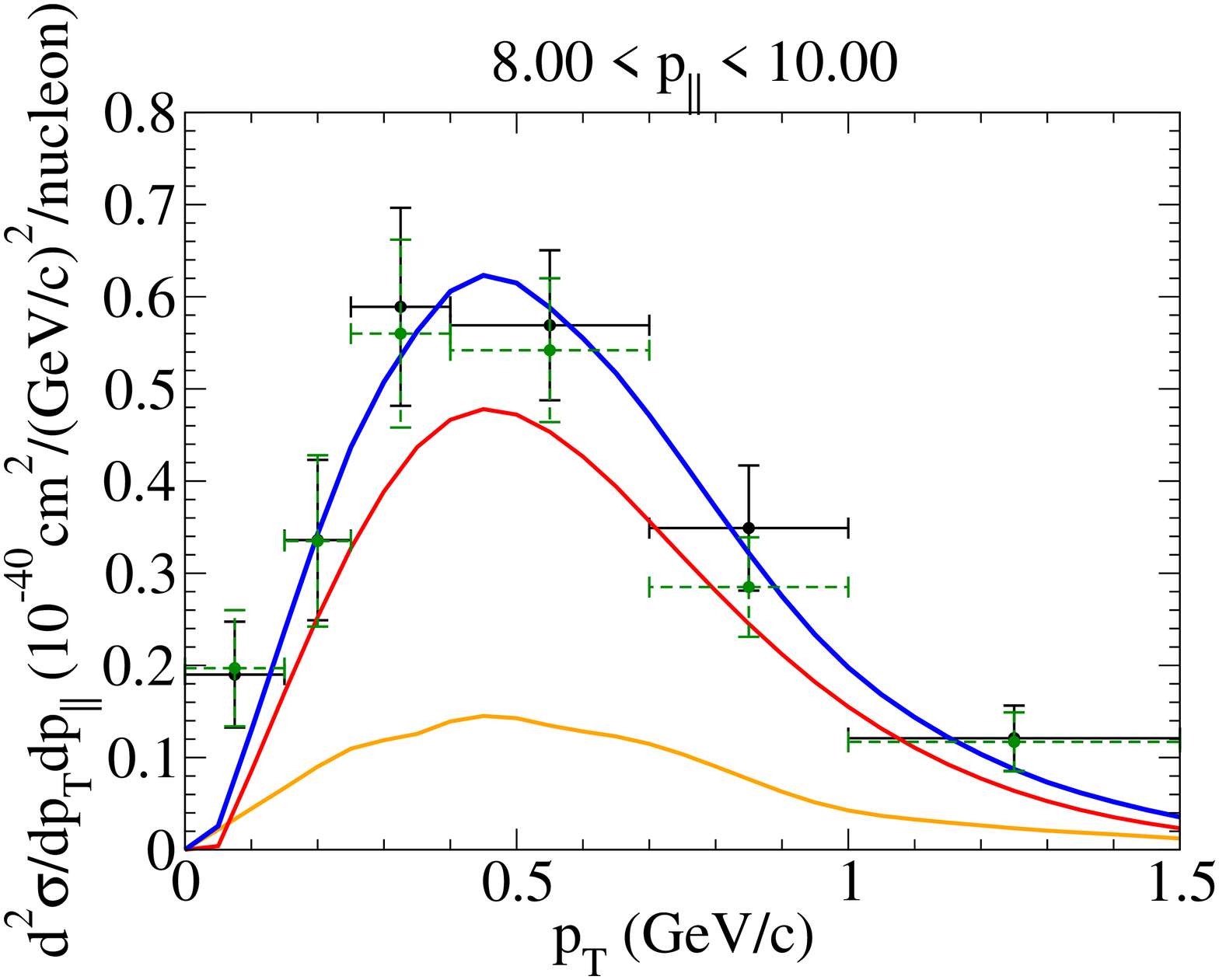}\hspace*{-0.584cm}%
		\hspace*{-0.295cm}\includegraphics[scale=0.192, angle=270]{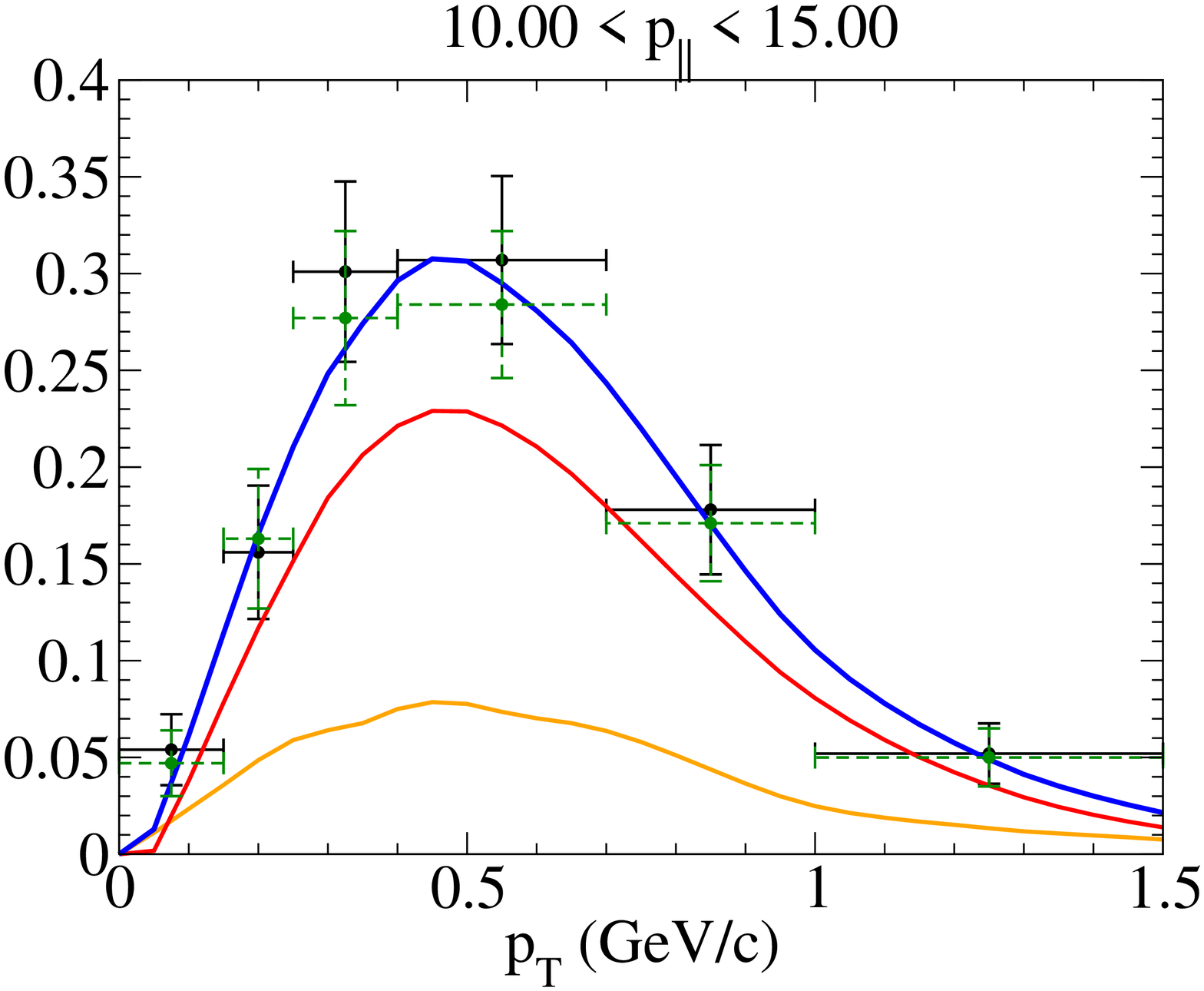}\hspace*{-0.584cm}%
	\caption{(Color online) The MINERvA ``QE-like" and ``CCQE" double differential cross sections for $\bar\nu_\mu$  scattering on hydrocarbon versus the muon transverse momentum, in bins of the muon longitudinal momentum (in GeV/c). The curves represent the prediction of the SuSAv2+2p2h-MEC (blue) as well as the separate quasielastic (red) and 2p2h-MEC (orange) contributions.  The data and the experimental antineutrino flux are from Ref.~\cite{Patrick:2018gvi}}.
	\label{fig:fig1}
\end{figure*}

It can be seen that the agreement with the data is good in all cases, and only a few data points are slightly underestimated by our calculation. Likewise, these ``CCQE'' data  have larger model-dependent systematic uncertainties than the ``QE-like'' ones due to increased reliance on the GENIE resonance and FSI models for background subtraction.  In Fig.~\ref{fig:fig1} the separate pure QE and 2p2h-MEC contributions are also shown: the MEC are sizeable at all kinematics and they are essential in order to reproduce the data. 

\begin{figure*}[!h]\vspace*{0.24cm}
	\begin{center}
		\hspace*{-0.295cm}\includegraphics[scale=0.192, angle=270]{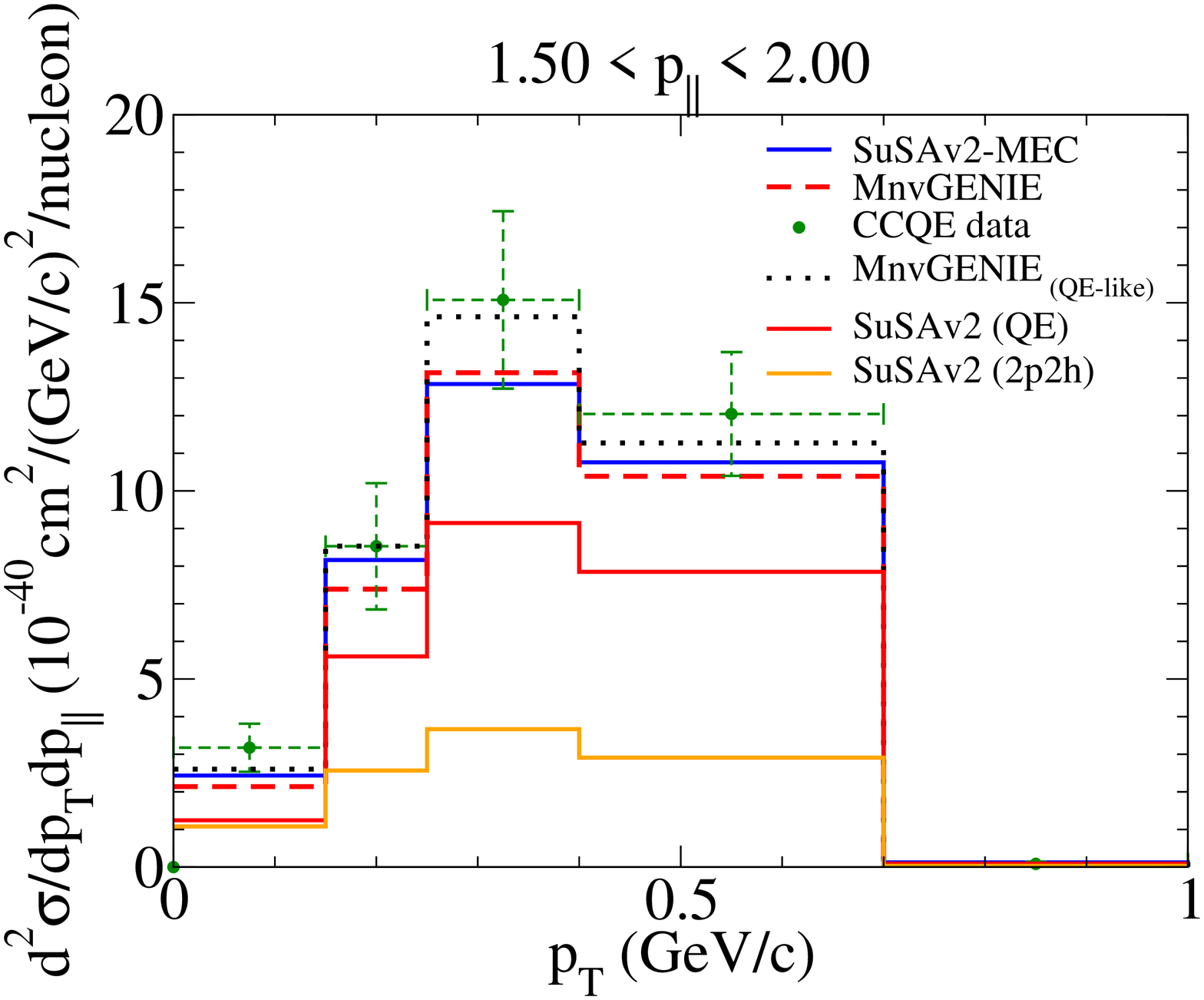}
		\hspace*{-0.65cm}\includegraphics[scale=0.192, angle=270]{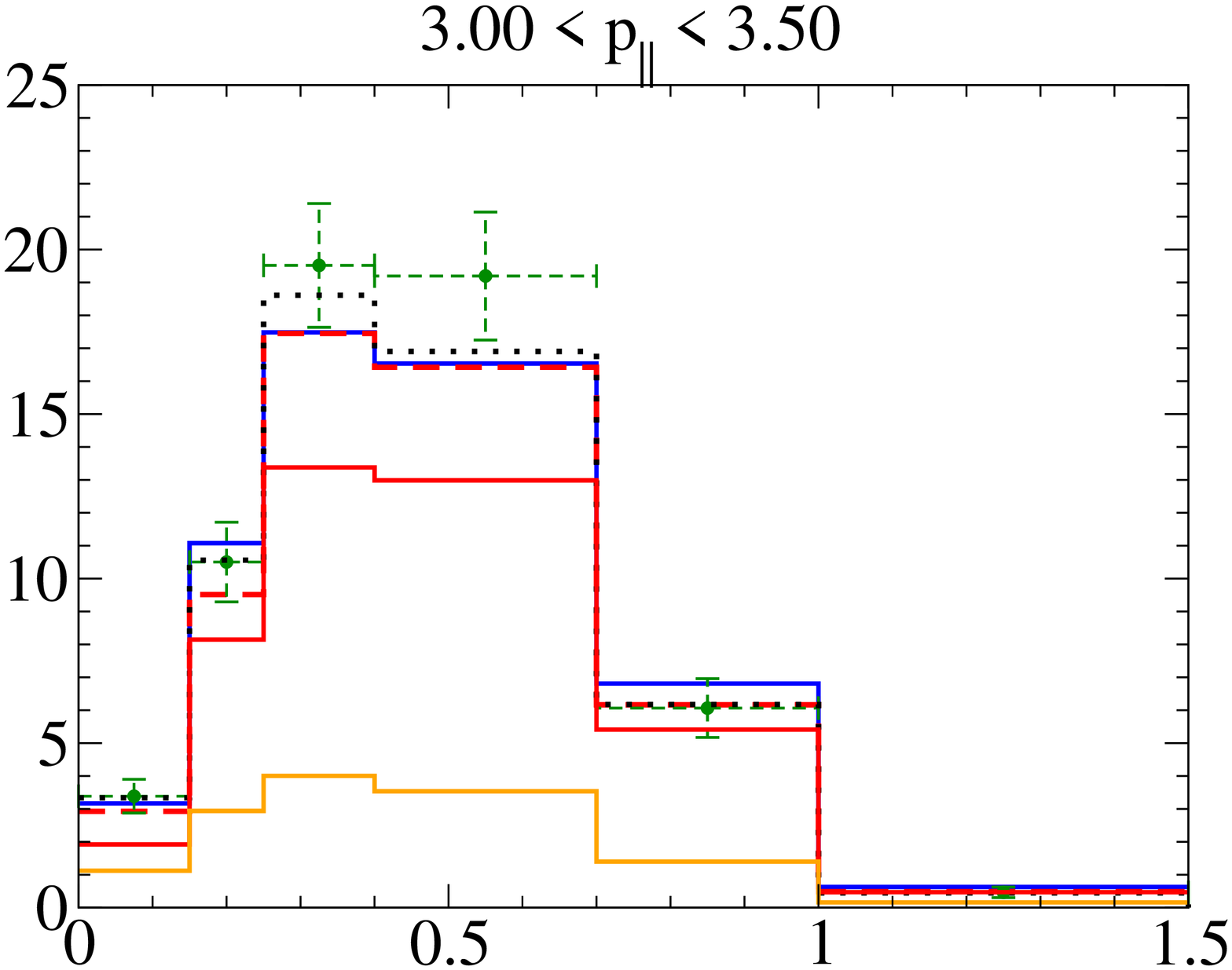}
		\hspace*{-0.65cm}\includegraphics[scale=0.192, angle=270]{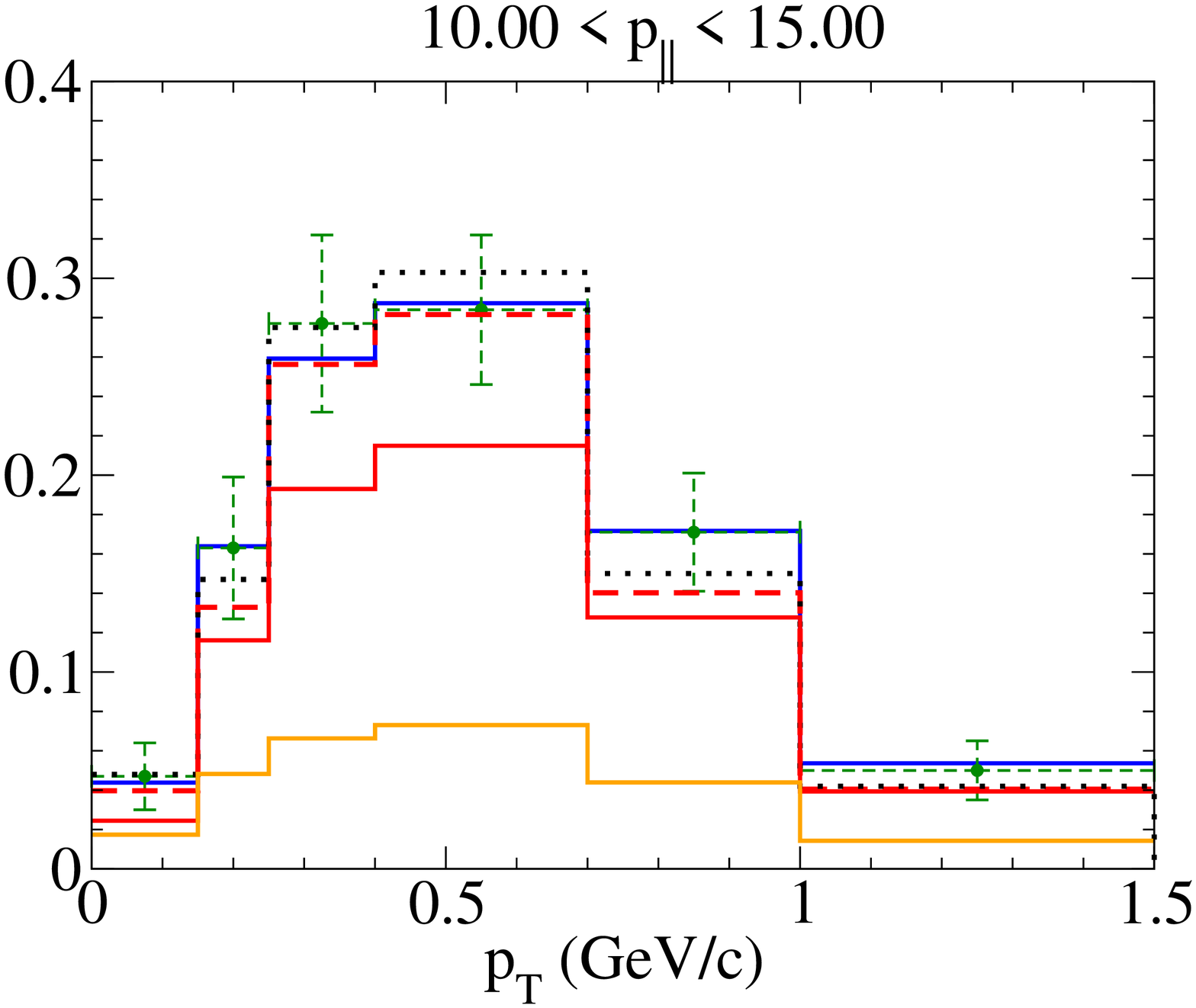}\\
		\hspace*{-0.295cm}\includegraphics[scale=0.192, angle=270]{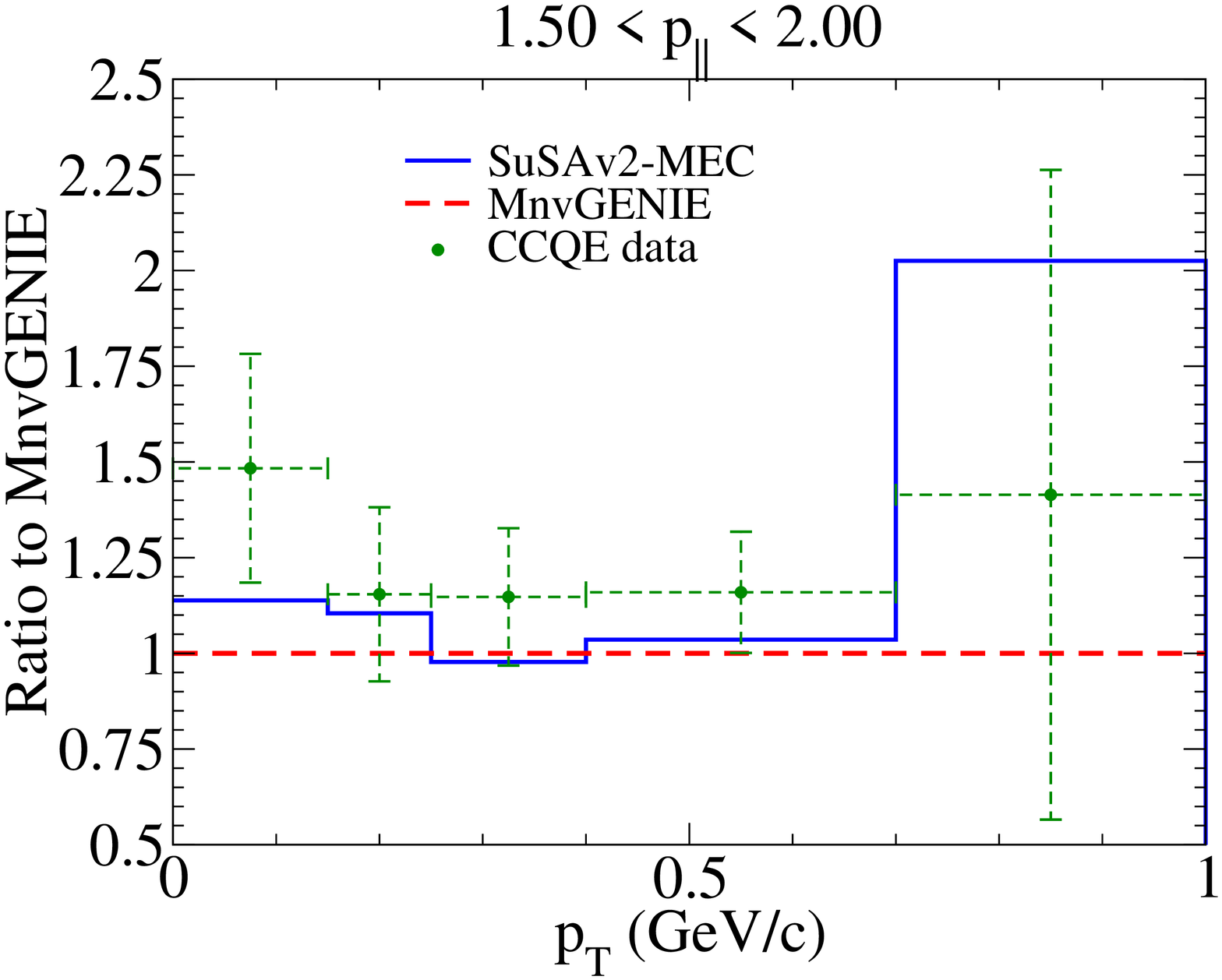}
		\hspace*{-0.65cm}\includegraphics[scale=0.192, angle=270]{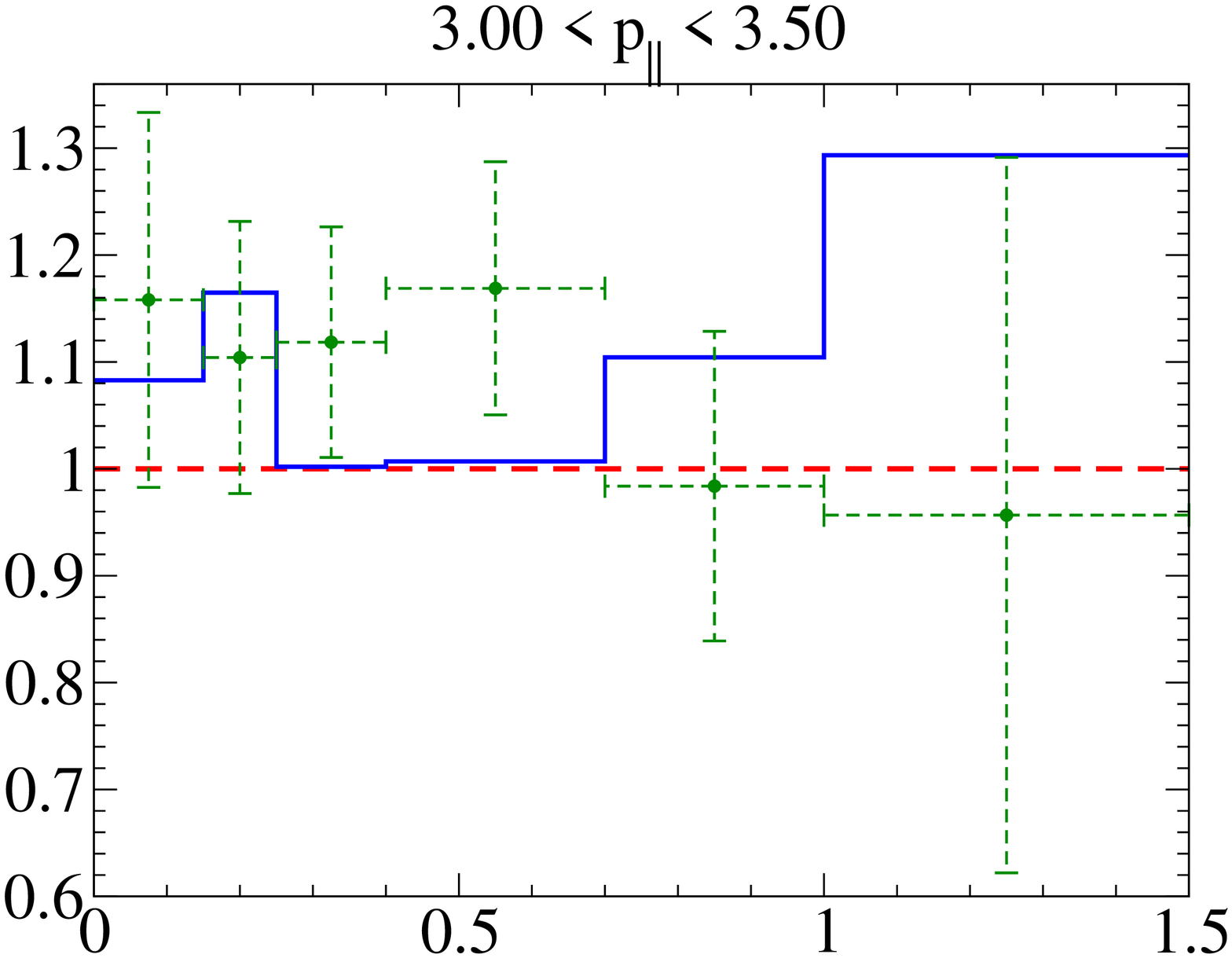}
		\hspace*{-0.65cm}\includegraphics[scale=0.192, angle=270]{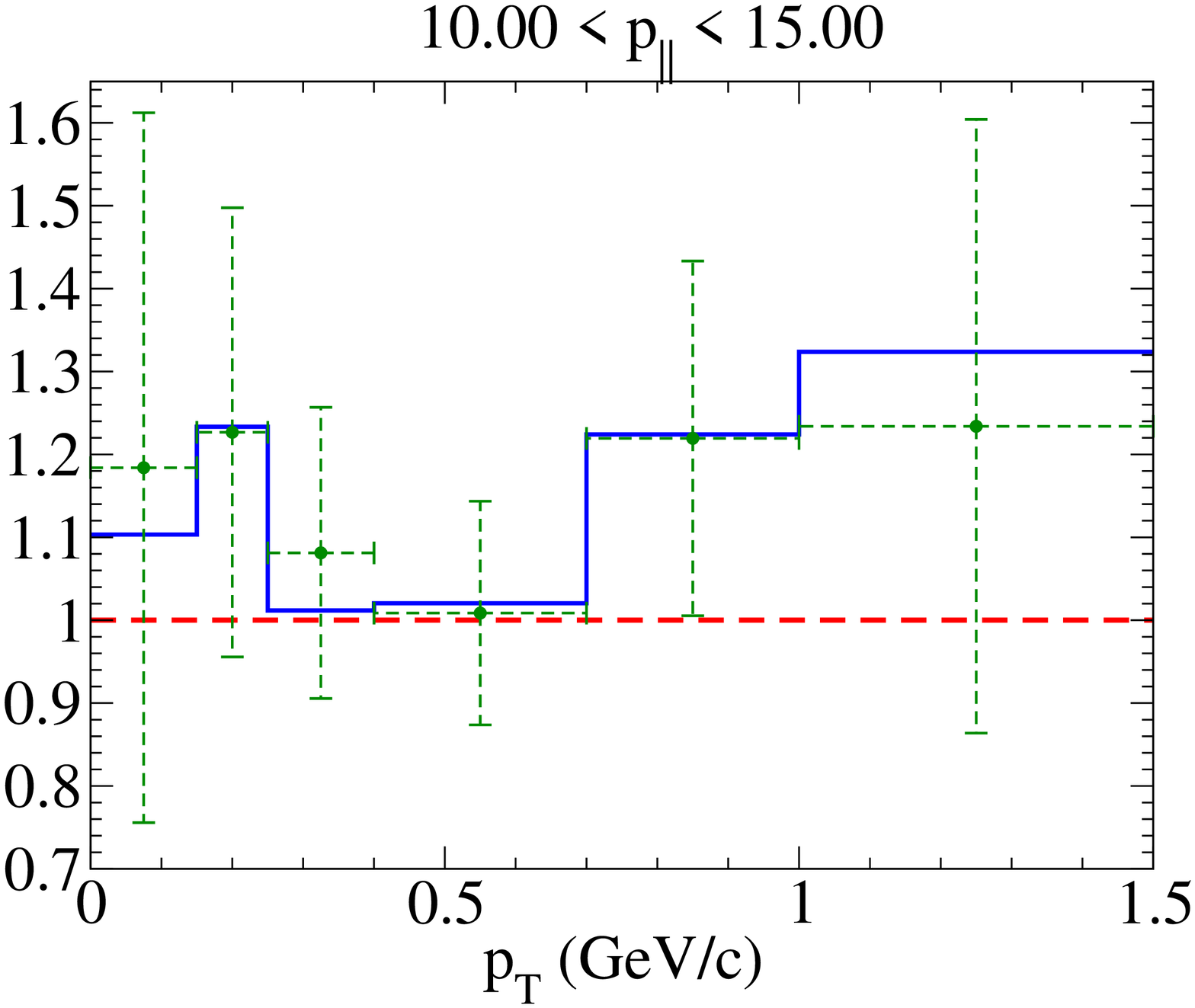}
	\end{center}\vspace*{-0.45cm}
	\caption{(Color online) Top panels: Comparison of SuSAv2-MEC with MINERvA ``CCQE'' data and MINERvA-tuned GENIE model for $\bar\nu_\mu$  scattering on hydrocarbon versus the muon transverse momentum, in bins of the muon longitudinal momentum (in GeV/c). The curves represent the prediction of the SuSAv2+2p2h-MEC (blue) as well as the ``CCQE'' MnvGENIE ones (red dashed). For completeness, we also show the ``QE-like'' MnvGENIE results as well as the separate QE and 2p2h channels in the SuSAv2-MEC model. Bottom panels: As top panels but showing ratio of data and SuSAv2-MEC predictions to GENIE. The comparison is quantified and expressed as $\chi^2$ per degree of freedom (d.o.f.) considering all bins, being $\chi^2$/d.o.f=1.79 for SuSAv2-MEC and $\chi^2$/d.o.f=1.58 for GENIE.} 
\label{fig:fig3}
\end{figure*}

A detailed comparison between the SuSAv2-MEC prediction and the extracted ``CCQE'' cross section as a ratio to the tuned GENIE prediction from the experimenters' data release (MINERvA-tuned GENIE~\cite{Rodrigues:2015hik,Gran:2018}) is shown in Fig. 2. MINERvA-tuned GENIE (MnvGENIE) is a modified version of the GENIE event generator tuned to MINERvA inclusive neutrino scattering data that incorporates nuclear effects such as weak nuclear screening and two-particle, two-hole enhancements. At MINERvA kinematics, the SuSAv2-MEC results seem to be larger than the MnvGENIE ones at the extreme $p_T$-bins and very close to them at the central values of $p_T$.  Overall, the comparison with MINERvA data is not very different for the two models, as we have checked by performing a $\chi^2$ test using the data release from~\cite{Patrick:2018gvi}. This test allows us to estimate quantitatively the level of agreement between data and predictions, accounting for the significant correlations between the data points. Considering all bins, i.e. for 58 degrees of freedom (d.o.f.), we have obtained $\chi^2$/d.o.f=1.79 for SuSAv2-MEC and $\chi^2$/d.o.f=1.58 for MnvGENIE.  Thus, the $\chi^2$/d.o.f. values obtained using the SuSAv2-MEC model turn out to be compatible with the MnvGENIE ones and with MINERvA data. For completeness, the QE-like MnvGENIE results are also displayed in Fig.~\ref{fig:fig3} to show the relevance of the effect beyond the CCQE+2p2h regime at MINERvA kinematics. In general, QE-like MnvGENIE results are similar but a bit larger than the SuSAv2-MEC ones apart from the region of large $p_{||}$ and $p_T$ values where SuSAv2-MEC seems to produce slightly higher results than QE-like MnvGENIE. The reasons for the larger SuSAv2-MEC results with respect to MnvGENIE may be related to the relativistic treatment of the FSI in the SuSAv2(RMF) for the 1p1h sector which causes a long tail that extends beyond the QE peak region. This is in principle absent in the RFG model used for CCQE interactions in MnvGENIE.


It is also important to note that both the Fermi gas QE model and the IFIC Valencia 2p2h model~\cite{Nieves:2011pp,Nieves:2011yp,Gran:2013kda} implemented in the tuned MnvGENIE simulation have been manually augmented to overcome the shortcomings of the simple QE model employed. This is done through the application of an RPA screening, extracted from~\cite{Nieves:2004wx}, to the GENIE QE model. Also the 2p2h model is empirically enhanced to describe the dip region of MINERvA's neutrino data in \cite{Rodrigues:2015hik}, before being applied to the antineutrino model~\cite{Gran:2018} and then used to compare to the measurement in~\cite{Patrick:2018gvi}. Such an empirical tune to the 2p2h component simultaneously accounts for dip-region shortcomings in the QE, 2p2h-MEC, and resonance models of GENIE. Conversely, the SUSAv2 model, based on RMF predictions, naturally puts additional QE cross section strength in the dip region and also has a stronger dip-region 2p2h-MEC component compared to the one used by MnvGENIE without resorting to any empirical tuning.

\begin{figure*}[!h]\vspace*{0.24cm}
	\begin{center}
		\hspace*{-0.295cm}\includegraphics[scale=0.192, angle=270]{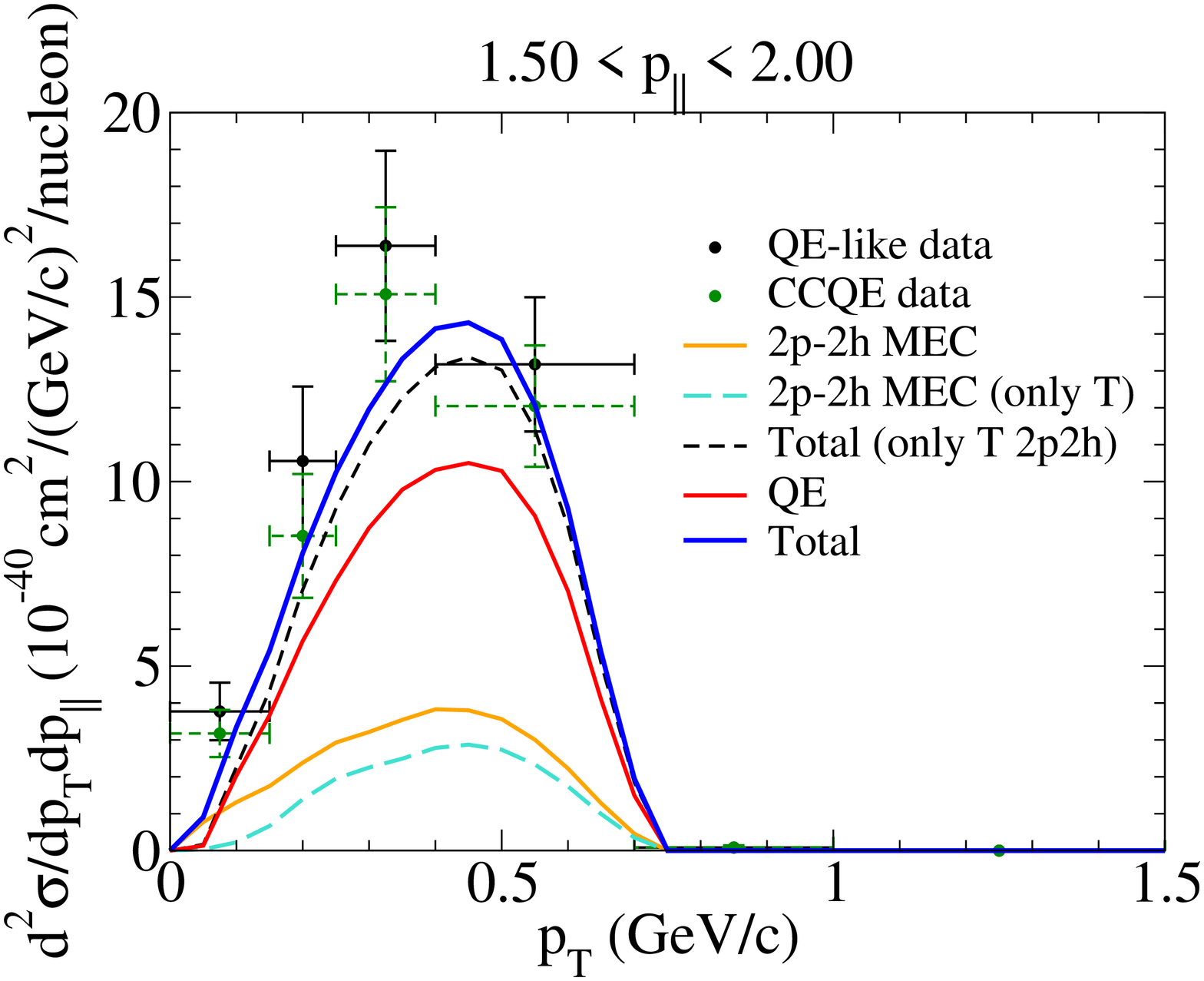}
		\hspace*{-0.65cm}\includegraphics[scale=0.192, angle=270]{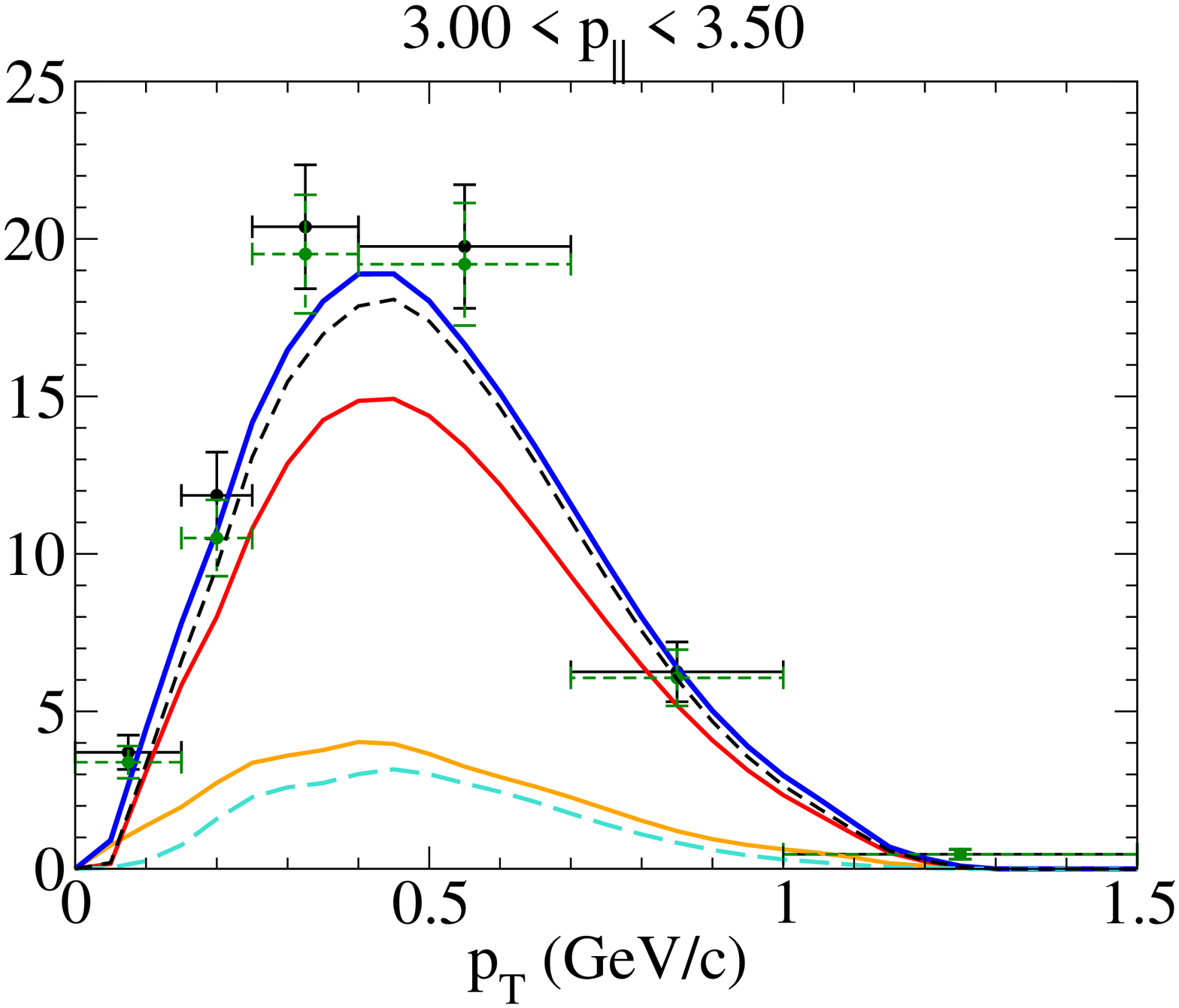}
		\hspace*{-0.65cm}\includegraphics[scale=0.192, angle=270]{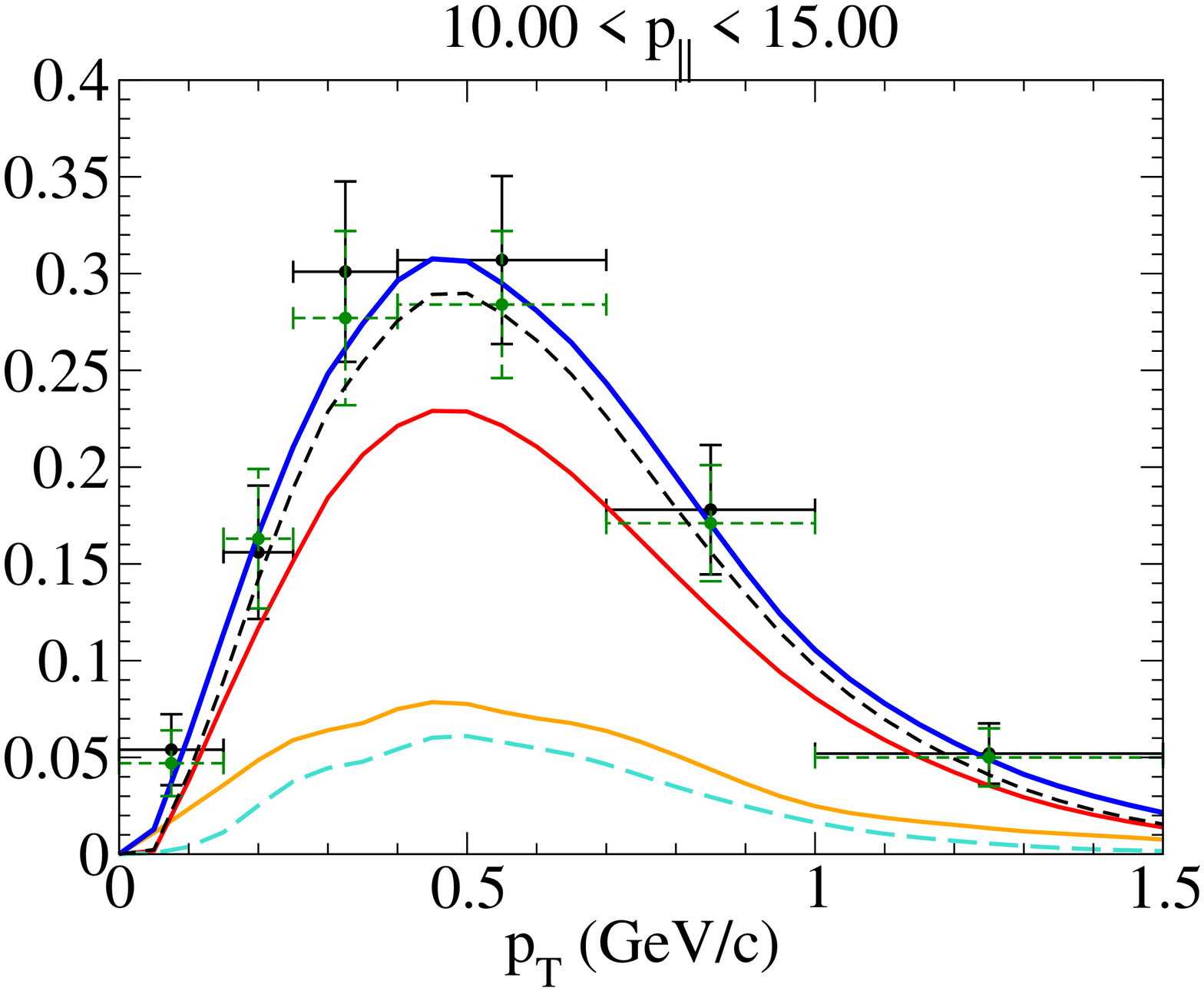}
	\end{center}\vspace*{-0.45cm}
	\caption{(Color online) As Fig.1, but showing the separate contribution of the pure transverse MEC (dashed curves) to also stress the relevance of the longitudinal MEC channel. } 
\label{fig:fig2}
\end{figure*}

A closer investigation of the MEC contribution is presented in Fig.~\ref{fig:fig2}, where we illustrate the role of the longitudinal MEC 
\footnote{Here, using the standard terminology of electron scattering studies, ``longitudinal" refers to the direction of the momentum transfer $\vec q$.} for three kinematics, including those corresponding to the lowest and highest bins of $p_L$. 
The result including the full -- longitudinal and transverse -- MEC contribution (solid lines) is compared to the one obtained using the pure transverse currents (dashed lines), resulting in a difference of around $\sim30-35\%$ in the 2p2h channel.
It appears that the longitudinal two-body currents, sometimes neglected in phenomenological 2p2h models~\cite{Mosel:MEC,Martini:2012mec,Bodek:2011mec}, give a non-negligible contribution to the total cross section ($\sim 5-10\%$), which improves the agreement with data. Note that this is not true in the case of electron scattering, where the longitudinal, purely vector, MEC are indeed negligible. In the case of weak processes, however, they are significant in the axial channel, representing up to a $\sim30\%$ of the axial contribution, and they are particularly important for antineutrino reactions due to the destructive interference between the vector and axial currents. A more detailed discussion of this point can be found in~\cite{Megias:2014qva}. 

\begin{figure*}[!h]\vspace*{0.24cm}
	\begin{center}
		\hspace*{-0.295cm}\includegraphics[scale=0.192, angle=270]{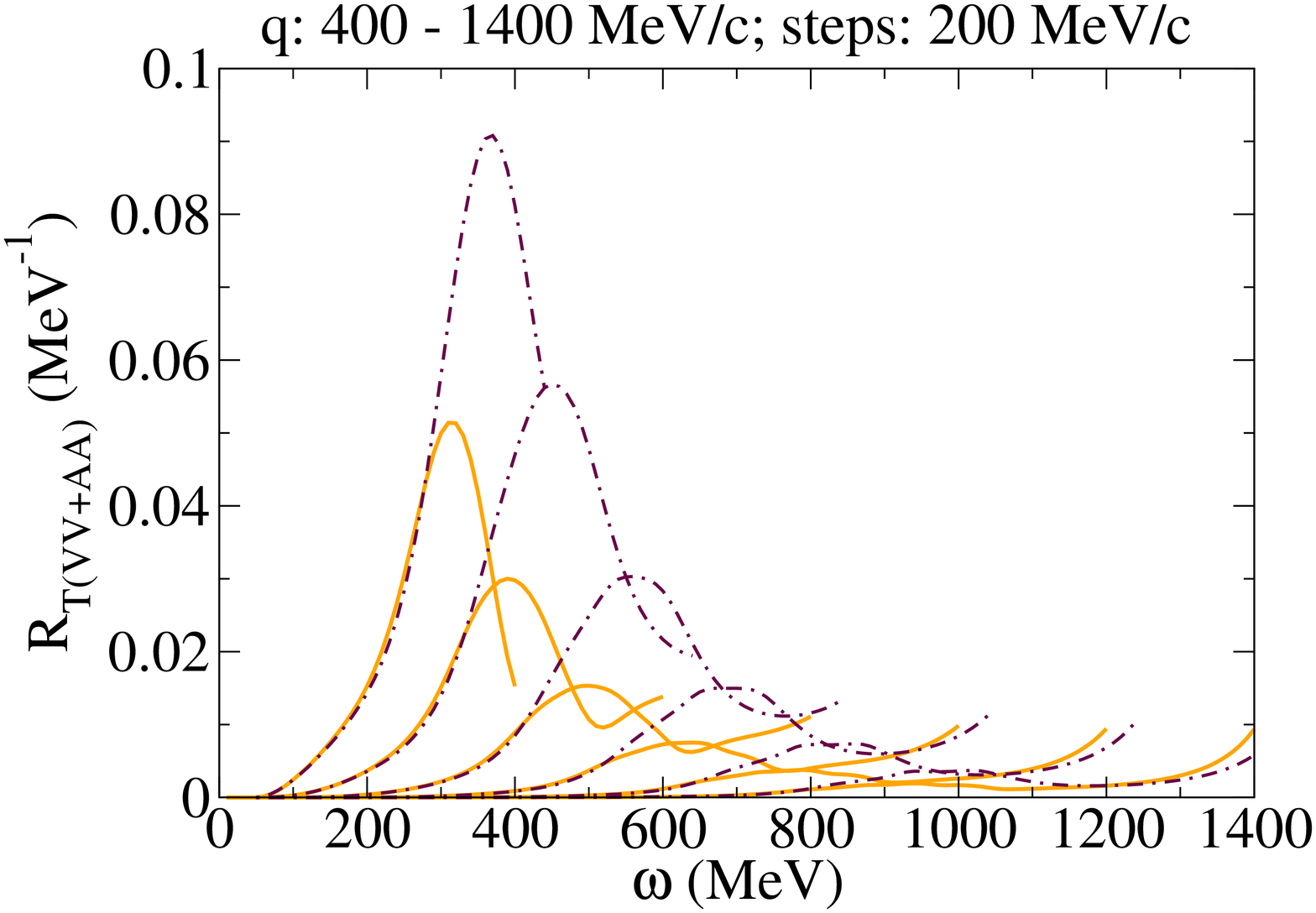}
		\hspace*{-0.15cm}\includegraphics[scale=0.192, angle=270]{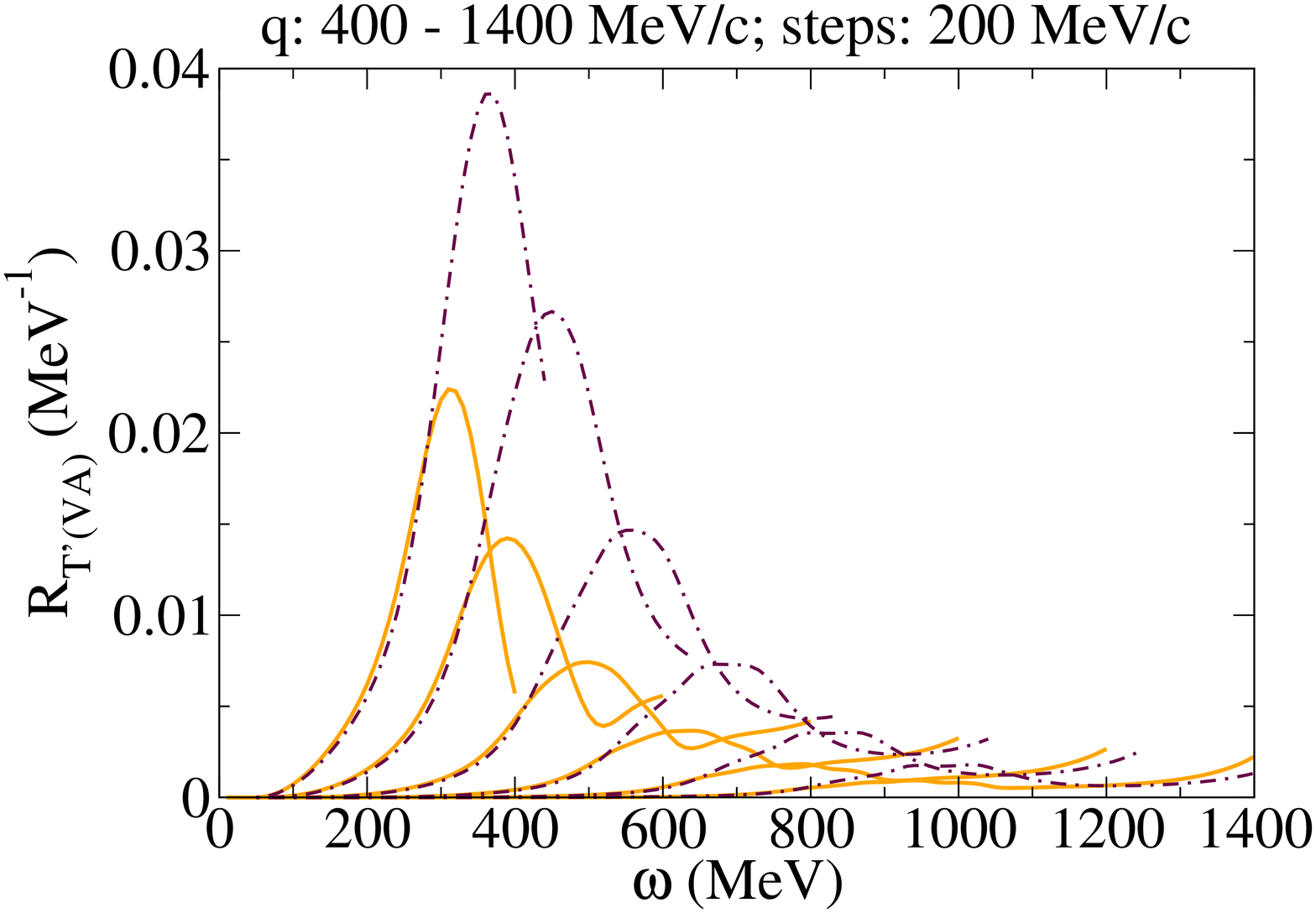}\\
		\hspace*{-0.295cm}\includegraphics[scale=0.192, angle=270]{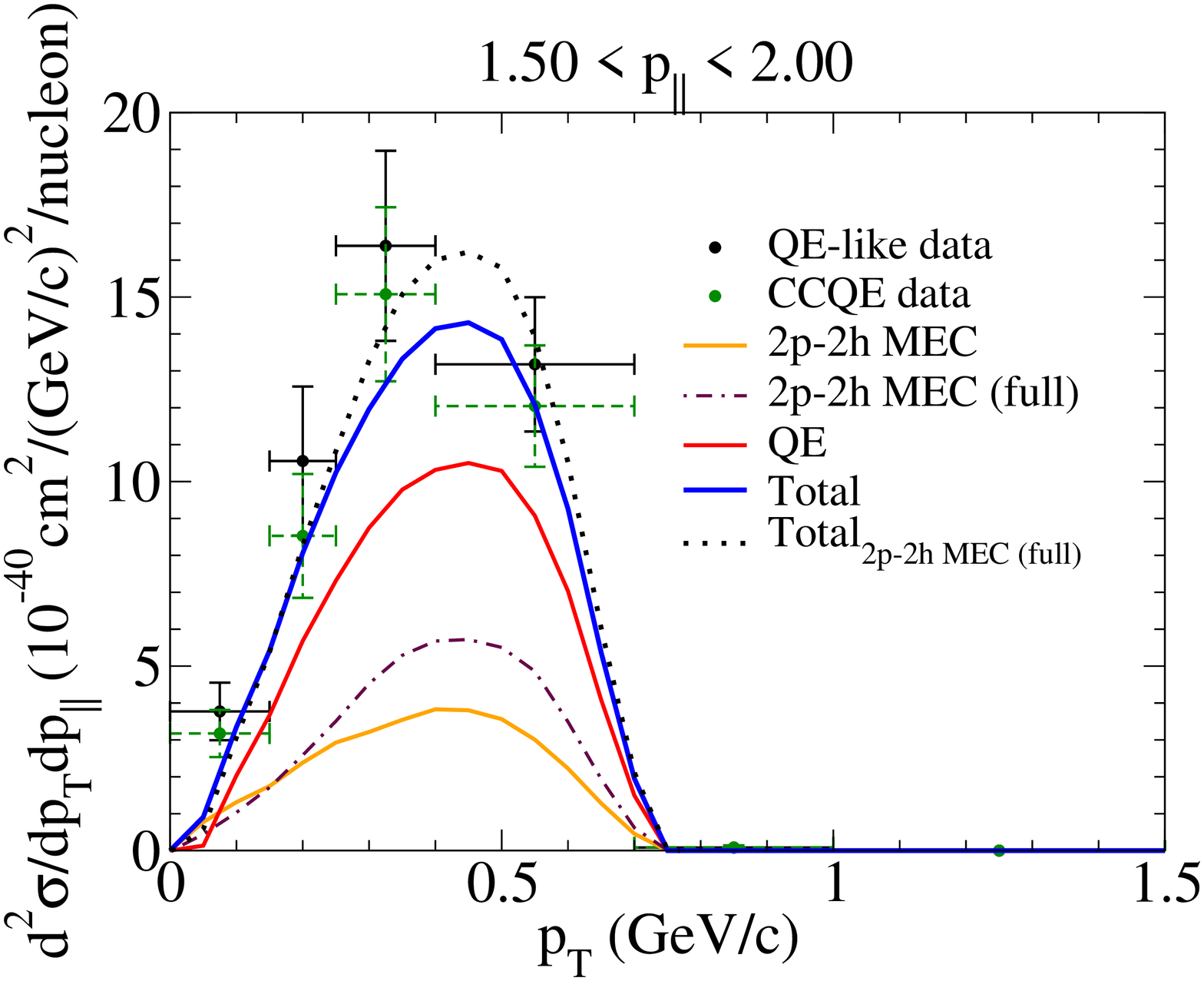}
		\hspace*{-0.65cm}\includegraphics[scale=0.192, angle=270]{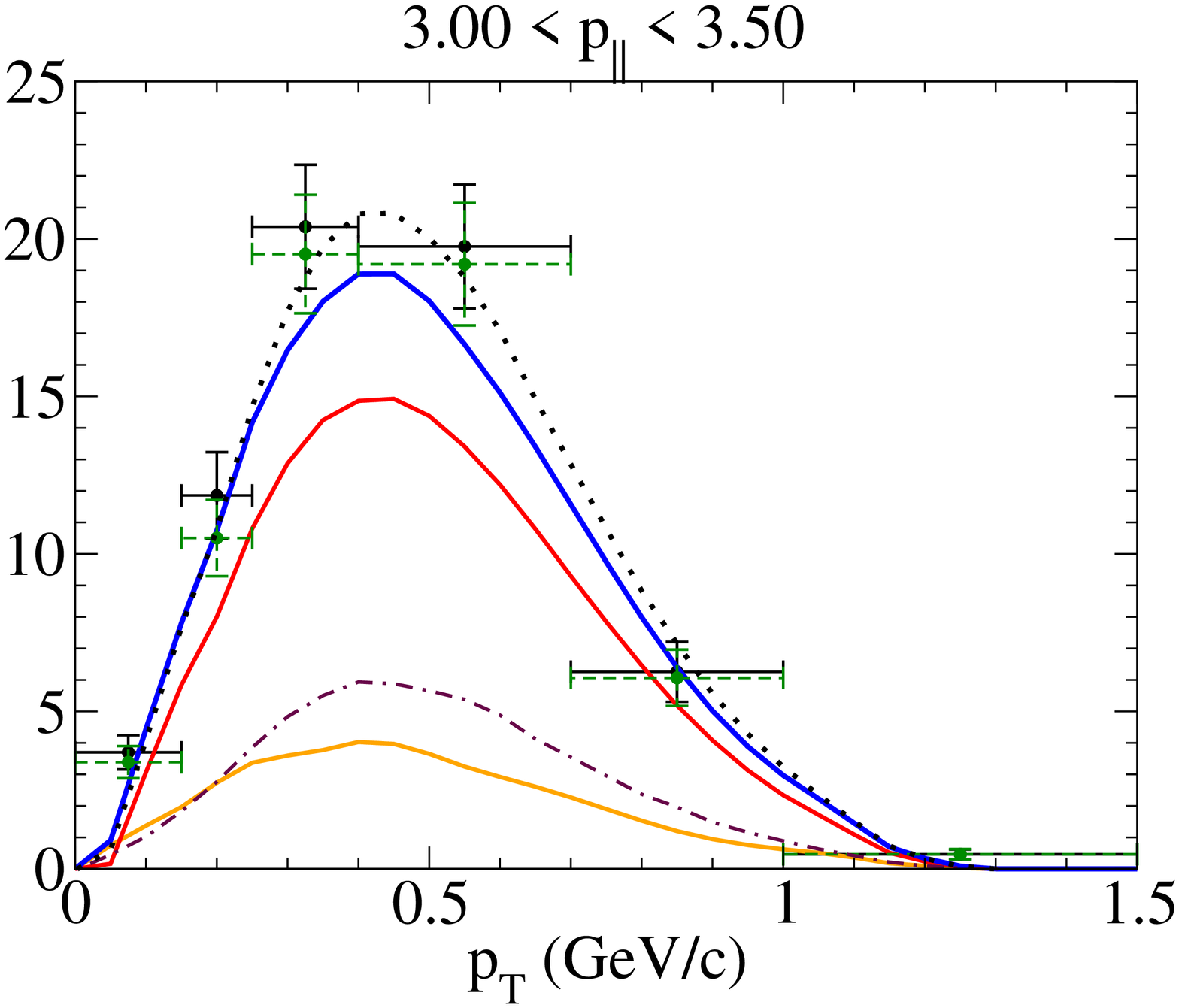}
		\hspace*{-0.65cm}\includegraphics[scale=0.192, angle=270]{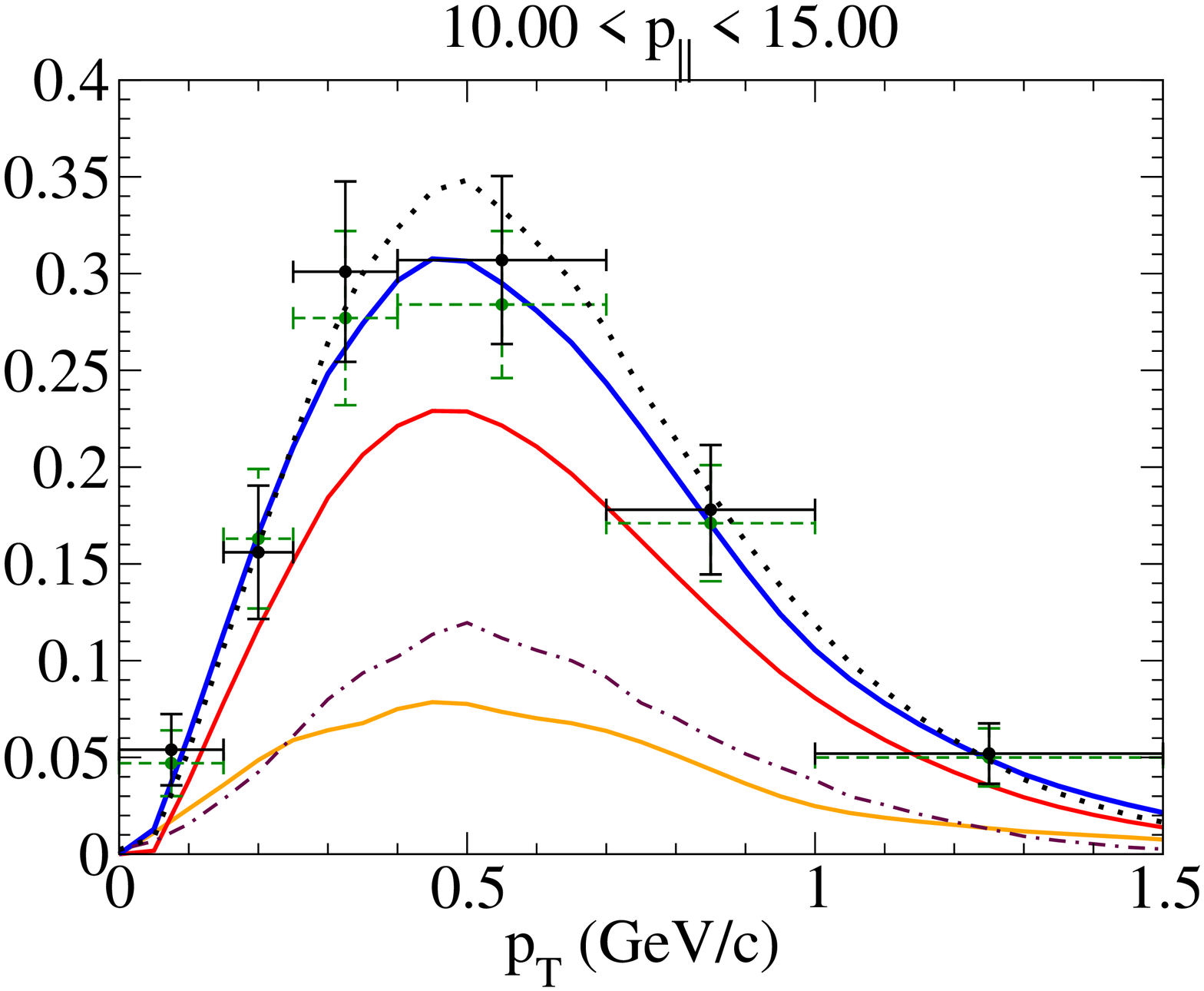}\\
	\end{center}\vspace*{-0.45cm}
	\caption{(Color online) Top panels: Comparison between purely real (solid lines) and full complex (dot-dashed lines)  $\Delta$ propagators for the 2p2h $R_T$ and $R_{T'}$ nuclear responses at different $q$ values. Bottom panels:
As Fig.1, but showing also the cross sections obtained by using the full complex $\Delta$ propagator in the 2p2h channel.}
\label{fig:fig4}
\end{figure*}

Before concluding, some comments are in order concerning the treatment of the $\Delta$-resonance propagator which appears in the two-body currents used to evaluate the 2p2h responses. In Figs.~\ref{fig:fig1}-\ref{fig:fig2}, following refs.~\cite{DePace:2003,Simo:2016ikv}, we have considered only the real part of
the $\Delta$ propagator. This prescription, also used by other groups~\cite{Rocco,DePace:2003,DePace:2004, Butkevich, Dekker, Alberico, VanOrden}, 
can be viewed as an empirical approach that leads to very good agreement with electron scattering data~\cite{Megias:2016lke}. In 
principle there are contributions stemming from the imaginary part of the Delta propagator which do not lead to pions in the final state.
These are dominated by a single diagram with a particle-hole self-energy insertion in the $\Delta$-hole Lindhard function that involves the square of the $\Delta$-propagator. The contribution of this diagram is large 
because it is the remnant of a double pole, a problem which affects all the second-order self-energy insertions of this kind (see, for instance,~\cite{Amaro:2010}).
This issue could in principle be solved by summing up the full series of self-energy insertions and employing everywhere the resummed $\Delta$ propagator. This, however, opens new problems, since the $\Delta$ state is being treated by adding an {\it ad hoc} width term and its extension to all orders is not trivial.
This latter task goes far beyond the scope of this paper. A more dedicated analysis of the effect and appropriateness of the full $\Delta$ propagator in ($e,e'$) and neutrino reactions will be addressed in forthcoming works.
Nevertheless, in order to estimate the relevance of the imaginary part of the $\Delta$ propagator, we present in Fig.~\ref{fig:fig4} the 2p-2h contributions using both real and full propagators for the analysis of the results shown in Fig.~\ref{fig:fig1}. 
The 2p2h contributions with the full propagator have been implemented using the microscopic model of~\cite{Simo:2016ikv} and, in order to optimize and speed up our calculations, we have employed the interpolation methods available in GENIE~\cite{Andreopoulos:2009rq}. 

As observed in the top panels of Fig.~\ref{fig:fig4}, the 2p2h nuclear responses obtained using the full propagator are around twice the ones where only the real part of the propagator is kept, the differences being negligible at low $\omega$ values with respect to $q$ ($\sim\omega<q/2$) and larger as $\omega$ increases. This observation, combined with the limited MINERvA's acceptance ($\theta_\mu<20\degree$, corresponding to $\cos\theta_\mu\gtrsim0.94$), which makes the low-intermediate $\omega$ and $q$ regions predominant, explains why  the 2p-2h cross sections using the full propagator, shown in the bottom panels of Fig.~\ref{fig:fig4}, do not differ largely ($\sim30-40\%$) from the ones obtained with the real propagator. The largest differences occur indeed at intermediate-high $p_T$ values, which are mostly related to large $E_\nu$ values, i.e. larger $\omega$ and $q$ values. Furthermore, the relatively small difference between the two results is also connected to the fact that the relevant $T'$ contribution is negative in antineutrino reactions and partly cancels the increase observed in the dominant $T$ responses when considering the full propagator. 
 In general, due to the large experimental error bars, the data comparison can be considered acceptable ($\chi^2$/d.o.f.=2.30), although the results with the full propagator tend to overestimate slightly the CCQE data at intermediate-high $p_T$ values.

In spite of the reasonable agreement with data of the SuSAv2-MEC with both real and full propagators, it is 
very important to point out that the application of the full propagator for the analysis of ($e,e'$) data would lead to a relevant increase of the cross section in the region of the 
$\Delta$-peak 
thus implying an 
overestimation of electron scattering data when combined to pion electroproduction models which describe well the data in the $\Delta$ peak~\cite{Megias:2016lke,Megias:2016fjk,Ivanov:2015aya,Rocco:2015cil}.
\newline

All results shown in this work correspond to the use of the commonly employed dipole axial nucleon form factor with the axial mass fixed to its standard value, $M_A=1.032$  GeV. The sensitivity of the SuSAv2 model to the description of the nucleon form factors and different choices of the parameters in the scaling function has been analyzed in detail in \cite{Megias:2016lke,Megias:2016fjk,Megias:2017PhD}, showing very tiny effects. Dependence of the SuSAv2 model including 2p2h-MEC (denoted as SuSAv2-MEC) on the Fermi momentum and shift energy has been also studied at depth in \cite{Megias:2017cuh,Amaro:2017den,Barbaro:2018arfg}, proving the robustness of the model to describe successfully (anti)neutrino cross sections for different nuclear targets. Furthermore, the SuSAv2-MEC model, which translates demanding microscopic calculations into a relatively straightforward formalism, can be easily implemented into MonteCarlo event generators as was done for \cite{Gran:2013kda,Wilkinson:2016wmz,Schwehr2017}
used in GENIE~\cite{Andreopoulos:2009rq}, NEUT~\cite{Hayato:2009zz}, and NuWro~\cite{PhysRevC.86.015505}. As with other microscopic calculations implemented in these event generators, SuSAv2 can be added as an initially inclusive calculation but extended to semi-inclusive predictions using an assumed initial nucleon momentum distribution and a semi-classical FSI cascade.  This model can then be employed in the analysis of present and forthcoming neutrino experiments. Work along this line is presently in progress~\cite{Megias:2018genie}.

Summarizing, the SuSAv2-MEC has been shown to provide a good description of the double differential $\bar\nu_\mu$-$CH$ inclusive cross sections recently measured by the MINERvA experiment.
The role of meson exchange currents has been proved essential in order to describe the data and the contribution of the MEC longitudinal components has been shown to be non-negligible and to improve the agreement with the data.
The successful comparison with these inclusive data, which correspond to a beam energy of $\sim$ 3.5 GeV, reflects the importance of a fully relativistic treatment. Work is in progress to extend the model to semi-inclusive reactions and to inelastic channels, which are needed for the analysis of future high energy neutrino oscillation experiments aiming to make high precision measurements of the leptonic CP violating phase.   

\begin{acknowledgments}
This work was partially supported by the INFN under project
MANYBODY, by the University of Turin under contract BARM-RILO-17, by
the Spanish Ministerio de Economia y Competitividad and ERDF (European Regional Development
Fund) under contracts FIS2017-88410-P, by the Junta de
Andalucia (grant No. FQM160). MBB acknowledges support from the ``Emilie du Ch\^atelet" programme of the P2IO LabEx (ANR-10-LABX-0038).
GDM acknowledges support from a Junta de Andalucia fellowship (FQM7632, Proyectos de Excelencia 2011).
SD and GDM acknowledge the support of CEA, CNRS/IN2P3 and P2IO.
We thank Richard Gran (MINERvA Collaboration) for careful reading of the manuscript and very helpful suggestions and  Arturo De Pace for enlightening discussions on 2p2h calculations. 
\end{acknowledgments}

\vspace*{-0.4cm}

\clearpage
\bibliography{biblio}

\end{document}